\documentclass[preprintnumbers,superscriptaddress,nofootinbib,aps,prd,floatfix,twocolumn]{revtex4}

\pdfoutput=1
\usepackage{graphicx}
\usepackage{latexsym}
\usepackage{amsfonts}
\usepackage{amssymb}
\usepackage{amsmath}
\usepackage{slashed}
\usepackage{feynmp}
\usepackage{hyperref}
\usepackage{url}
\usepackage{color}
\usepackage{cancel}
\usepackage{multirow,array}
\usepackage{float}
\usepackage{subfigure}
\usepackage{relsize}

\usepackage{lipsum}
\usepackage{amsmath, amssymb, amscd, amsthm, amsfonts}
\usepackage{graphicx}

\begin{document}
\date{\today}
\title{Constraining Dark Matter Substructure With {\it Gaia} Wide Binaries}

\author{Edward D.~Ramirez}
\affiliation{Department of Physics and Astronomy, Rutgers University, Piscataway, NJ 08854, USA}
\author{Matthew R.~Buckley}
\affiliation{Department of Physics and Astronomy, Rutgers University, Piscataway, NJ 08854, USA}

\begin{abstract}
We use a catalogue of stellar binaries with wide separations (up to $1$~pc) identified by the {\it Gaia} satellite to constrain the presence of extended substructure within the Milky Way galaxy. Heating of the binaries through repeated encounters with substructure results in a characteristic distribution of binary separations, allowing constraints to be placed independent of the formation mechanism of wide binaries.
Across a wide range of subhalo density profiles, we show that subhalos with masses $\gtrsim 65\,M_\odot$ and characteristic length scales similar to the separation of these wide binaries cannot make up 100\% of the Galaxy's dark matter. Constraints weaken for subhalos with larger length scales and are dependent on their density profiles. For such large subhalos, higher central densities lead to stronger constraints. Subhalos with density profiles similar to those expected from cold dark matter must be at least $\sim 5,000$ times denser than predicted by simulation to be constrained by the wide binary catalogue.

\end{abstract}


\maketitle

\section{Introduction} \label{sec:intro}

The particle nature of the dark matter remains an open question in physics. Measurements of the distribution, formation, and evolution of large-scale structure of the Universe are consistent with cold, collisionless dark matter, interacting with itself and with baryonic matter only through gravity, and seeded by primordial density fluctuations. This consistency between observation and predictions extends down to the scales of dwarf galaxies, $\sim 10^{8-9}\,M_\odot$ \cite{tegmark04}. Smaller objects are expected to exist but are difficult to directly observe, due to inefficient star formation in low-mass objects \cite{bullock2017, zavala2019}. Deviations from the predictions of gravity-only cold dark matter models on the structure or distribution of low-mass dark matter halos would be a sign of non-trivial physics within the dark sector, physics that may be difficult to probe any other way \cite{buckley2018}.

While gravitationally-bound dark matter halos below the mass of dwarf galaxies are expected to exist both independent of and as internal substructure to larger halos (in the latter case forming ``subhalos'' within the host), the lack of tracer stars within small halos
makes searching for such objects external to larger galaxies extremely difficult. Instead, constraints on low-mass halos tend to focus on subhalos within larger galaxies and galaxy clusters, which can reveal their presence through gravitational effects within the host object. Gaps observed in the Palomar 5 \cite{oden01} and GD-1 \cite{grillmair06} stellar streams \cite{banik2019_evidence, banik2018} within the Milky Way may be the result of substructure with mass $\sim10^{6}\,M_{\odot}$ \cite{banik2019_constraints}. Constraints on subhalos with masses down to $\sim 10^7\,M_\odot$ have been extracted from the measured flux ratios of quadruply-imaged quasars \cite{gilman2021, gilman2020, gilman2019, gilman2018}. Point-like dark matter substructure (primordial black holes \cite{carr_rev,green_2021} or massive compact halo objects---MACHOs \cite{griest_1993}) are also constrained over a wide range of masses by measurements of microlensing and tidal disruption \cite{carr2020}. However, it is important to note that the microlensing constraints assume the dark matter is highly compact, and in general do not apply if the dark matter is extended over scales larger than the Einstein radius of the microlensing event (which can be as small as ${\cal O}(10\,{\rm AU})$). 

In this paper, we develop a new probe of dark matter subhalos within the Milky Way using wide binary star systems (semimajor axes $\gtrsim 10^{-3} \ {\rm pc}$). While their component stars are on the main sequence, such binaries evolve as isolated two-body systems \cite{yoo04,longhitano2011}, unless tidal forces act on them. Subhalos can exert such forces by passing near a binary. During such fly-by encounters, the subhalos inject energy into the binary, causing the binary's semimajor axis to increase, and eventually resulting in complete disruption of the bound system \cite{banik2020}. This ``heating'' is more effective for more widely separated binaries, and so a population of perturbing subhalos acting on a population of binaries results in a rapid decrease in the number of binaries as a function of their projected separation on the sky \cite{yoo04,weinberg87,galtext}. The heating of wide binaries has been used to place limits on primordial black holes and other point-like perturbers within the Milky Way \cite{yoo04,quinn09,rodriguez14,tyler22}. In this paper, we apply the formalism to extended dark matter objects. Related constraints on point-like perturbers have been obtained using the heating of large stellar clusters within the Eridanus II dwarf galaxy \cite{brandt2016} and the Milky Way disk \cite{lacey85}, which may also potentially be used to constrain extended dark matter objects. A constraint based on the heating of binaries within dwarf galaxies has been proposed \cite{penarrubia2010,penarrubia2016}, though the necessary data is not yet available. 

Our constraints are set using a sample of 9,637 binaries, selected from a catalogue by Ref.~\cite{ebr21} (hereafter referred to as E21) that was constructed using data from the {\it Gaia} Early Data Release 3 (eDR3) \cite{gaia_21, gaia_16}. {\it Gaia}'s precision photometric and astrometric measurements allowed E21 to identify pairs of stars whose physical separation and relative velocities are consistent with bound Keplerian orbits \cite{ebr2018,tian19,ebr21,jimenez19,hartman20}. Our sample of binaries is consistent with membership in the Milky Way's stellar halo \cite{bahcall1980} and thick disk \cite{gilmore1983}. Compared to binaries in the thin disk, the higher ages and sparsity of baryonic sources in these regions of the Galaxy implies that their binaries are the most affected by dark matter substructure and the least affected by baryonic tidal perturbers. However, our constraints on subhalos make the conservative assumption that the observed present-day distribution of projected binary separations is due solely to dark matter subhalo encounters.

The paper is structured as follows: we present the sample of binaries in Section~\ref{sec:catalog}. In Section~\ref{sec:theory}, we model the heating of binaries as a result of tidal forces exerted by subhalos. In Section~\ref{sec:stats}, we develop the statistical methods used to set constraints on subhalos. In Section~\ref{sec:results}, we set constraints on subhalos with a wide variety of density profiles, including density profiles predicted by N-body simulations of cold dark matter. We make concluding remarks in Section~\ref{sec:conclusions}. 

\begin{figure}[t!]
    \centering
    \includegraphics[width = 0.9\columnwidth]{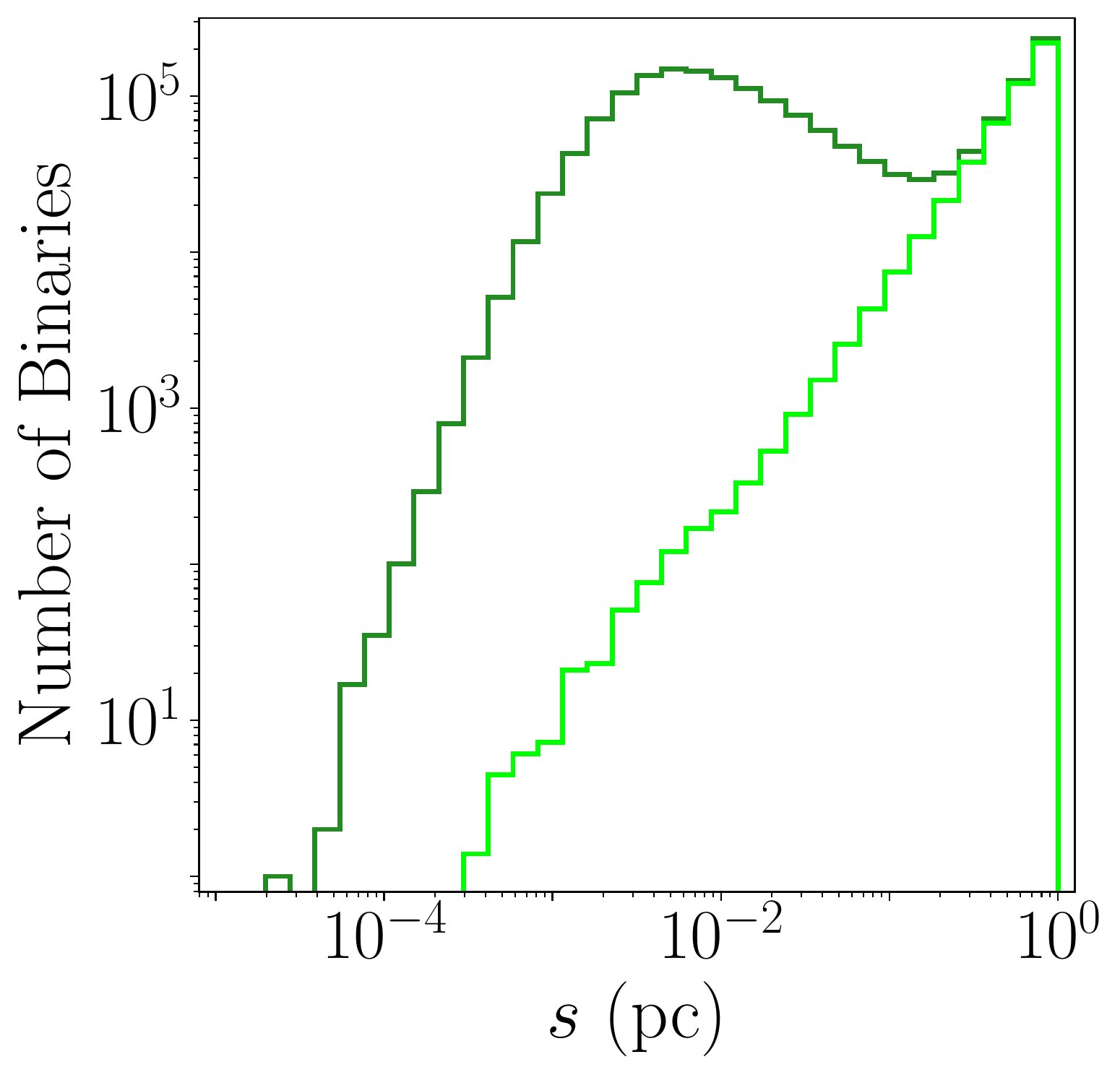}   
    \caption{Projected separation distribution of the initial binary catalogue \cite{ebr21} (dark line) and the estimated number of contaminants (bright line), obtained by weighting each binary by ${\rm min}(\mathcal{R},1)$ for contamination probability estimate $\mathcal{R}$.} \label{fig:data_uncut}
\end{figure}

\section{{\it Gaia} Wide Binaries} \label{sec:catalog}

\begin{figure*}[t!]
    \centering
    \includegraphics[width = 1.8\columnwidth]{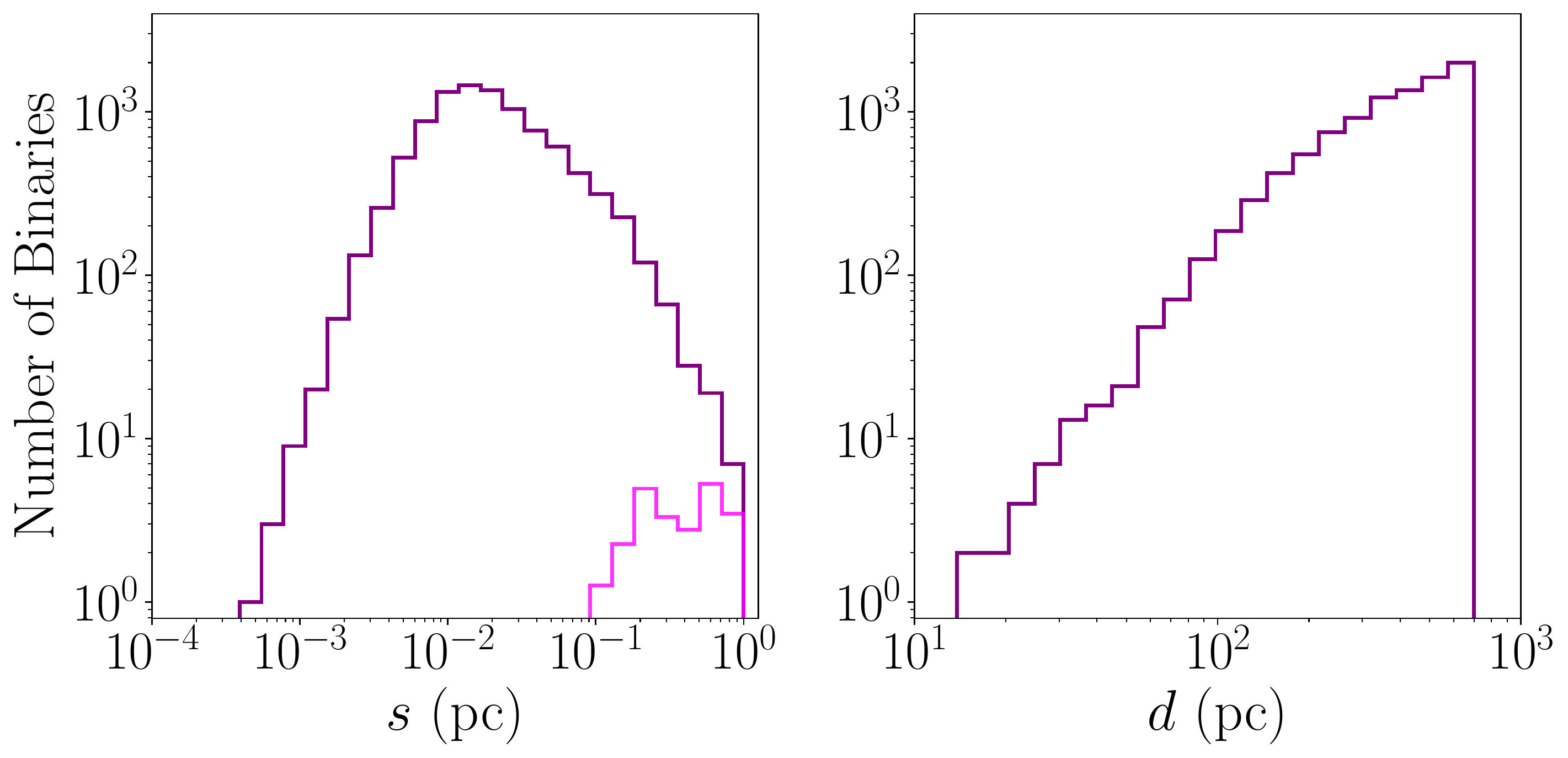}
    \caption{\textit{(Left)} Projected separation distribution of our selected sample of binaries (dark line), which we use to set constraints on subhalos. The bright line denotes the estimated number of contaminants, obtained by weighting each binary by its contamination probability $\mathcal{R}$. \textit{(Right)} Distribution of measured distances from Earth.} \label{fig:data}
\end{figure*}

Binaries with widely separated stellar components can offer strong constraints on a population of tidal perturbers, as the tidal force grows with the size of the system on which it acts. While a single encounter between a binary and a subhalo may not result in a significant change in the orbital parameters of the binary, multiple encounters over long timescales can slowly evolve the system to much larger separations or disrupt it altogether. 

Perturbations from baryonic sources (e.g., other stars, gas, and dust) lead to similar orbital evolution as perturbations sourced by dark matter substructure. While conservative limits on the population of dark matter perturbers can be set by assuming all the evolution is due to the dark sector, these constraints can be strengthened by using wide binaries that orbit in the stellar halo or the thick disk, where there are fewer baryonic sources. As the end-of-life evolution of a star off of the main sequence can introduce significant velocity kicks to binaries \cite{davis2008}---which would mimic and obscure the effect from perturber encounters---we further restrict ourselves to binaries whose component stars are long-lived main sequence stars.

We use for our dataset the collection of widely-separated binaries identified from within the {\it Gaia} Space Telescope's eDR3 \cite{gaia_21} by E21, using techniques first presented in Ref.~\cite{ebr2018}. Well-measured stars from {\it Gaia} eDR3 are grouped with their neighbors by identifying pairs whose measured relative velocities and separations are consistent with bound Keplerian orbits. Groups with three or more stars that appear bound are filtered out. This process results in an initial catalogue of 1,817,594 binary candidates, shown in Fig.~\ref{fig:data_uncut}. 

These binary candidates each have a projected separation $s$ (the distance between the component stars as projected onto the plane of the sky) ranging from $\sim 10^{-4}$~pc to $\sim 1$~pc. The low-separation tail of this distribution is set by decreasing sensitivity of the {\it Gaia} telescope at smaller angular separation \cite{2021A&A...649A...5F} and the difficulty of resolving overlapping stellar components with similar $G$-band magnitudes. To ensure our sample is complete at low separations, we use an empirical fitting function that describes the probability of \textit{Gaia} resolving stellar components with angular separation $\theta$ and $G$-band magnitude difference $\Delta G = |G_{1} - G_{2}|$ \cite{ebr2018}:
\begin{align}
    f_{\Delta G}(\theta) = \frac{1}{1 + (\theta/\theta_{0})^{-\beta}}, \label{eq:completeness_function}
\end{align}
where $\theta_{0}$ characterizes the angular separation below which \textit{Gaia} is insensitive to binaries and $\beta$ determines the rate at which \textit{Gaia}'s sensitivity drops to 0 for $\theta \ll \theta_{0}$. Following the approach of Ref.~\cite{ebr2018}, E21 fit the values of $\theta_{0}$ and $\beta$ for sources in a range of $\Delta G$ bins. We estimate these parameters for arbitrary $\Delta G$ by interpolating the fits over the binned data. Using this function, we select binaries with $f_{\Delta G} > 0.999$, which roughly corresponds to $\theta > 3 \ {\rm arcsec}$. 

Following Ref.~\cite{tian19} (hereafter referred to as T19), we select a subcatalogue of binary candidates each composed of two main sequence stars whose tangential velocities relative to the Sun are large, $v_\perp>85$~km/s, and whose distance from the Sun is less than 700~pc. Systems with such high tangential velocities are less likely to be members of the Milky Way's thin disk, and are instead likely members of the stellar halo or the thick disk \cite{chiba2000,bensby2003,venn04,Yoachim_2008}---both of which contain older stars \cite{kilic_2017,reid05, ivezic08} that have had fewer tidal interactions with baryonic perturbers as compared with stars in the thin disk. 

At large projected separation $s$, the rate of chance alignments (pairs of stars in the catalogue which are identified as binaries despite not being gravitationally bound in actuality) increases. The catalogue of E21 provides for each binary candidate an estimate of the probability $\mathcal{R}$ that the candidate is a chance alignment.\footnote{We emphasize that $\mathcal{R}$ is itself not strictly a probability, rather it is an estimate of a probability. Notably, in some cases, $\mathcal{R} > 1$.} The expected distribution of chance alignments in the initial catalogue, which can be estimated from these $\mathcal{R}$ values, is shown by the bright line in Fig.~\ref{fig:data_uncut}. In Section~\ref{sec:stats}, we will construct an empirical contamination model, treating $\mathcal{R}$ as the true contamination probability of binary candidates. We therefore select binaries with $\mathcal{R} \leq 1$. 

As in T19, we remove a majority of the chance alignments in the initial catalogue by selecting binaries satisfying $\Delta \mu \leq \Delta \mu_{\text {orbit }}+1.0 \sigma_{\Delta \mu}$ and $\sigma_{\Delta \mu} \leq 0.12$ mas yr$^{-1}$, where $\Delta \mu$ is the measured magnitude of the proper motion difference between stellar components, $\Delta \mu_{{\rm orbit}}$ is the maximum proper motion difference allowed if the components followed a circular orbit of total mass $5 \ M_{\odot}$, and $\sigma_{\Delta \mu}$ is the uncertainty in $\Delta \mu$. 

These selections leave us with a catalogue of 9,637 binary candidates, which we take as our sample. The distribution of projected separation $s$ is shown to the left of Fig.~\ref{fig:data}. Though the sample still has a low-separation tail, this is now mainly due to the incompleteness arising from the selection cut $f_{\Delta G} > 0.999$ rather than \textit{Gaia}'s sensitivity, making the incompleteness easier to model accurately (see Section~\ref{sec:stats}). Once this incompleteness is taken into account, the separation distribution can be fit by a broken power law breaking at $\sim 0.1 {\rm \ pc}$.\footnote{At high separations ($10^{-2} \ {\rm pc}$ to $1 \ {\rm pc}$), the data is nearly complete, and so the broken power-law behavior is most clearly seen here even without correcting for completeness.} As we will see in Section~\ref{sec:stats}, our limits will be set by this break. The distribution of distances from Earth, $d$, is shown to the right of Fig.~\ref{fig:data}.

\begin{figure*}[t!]
    \centering
    \includegraphics[width = 1.8\columnwidth]{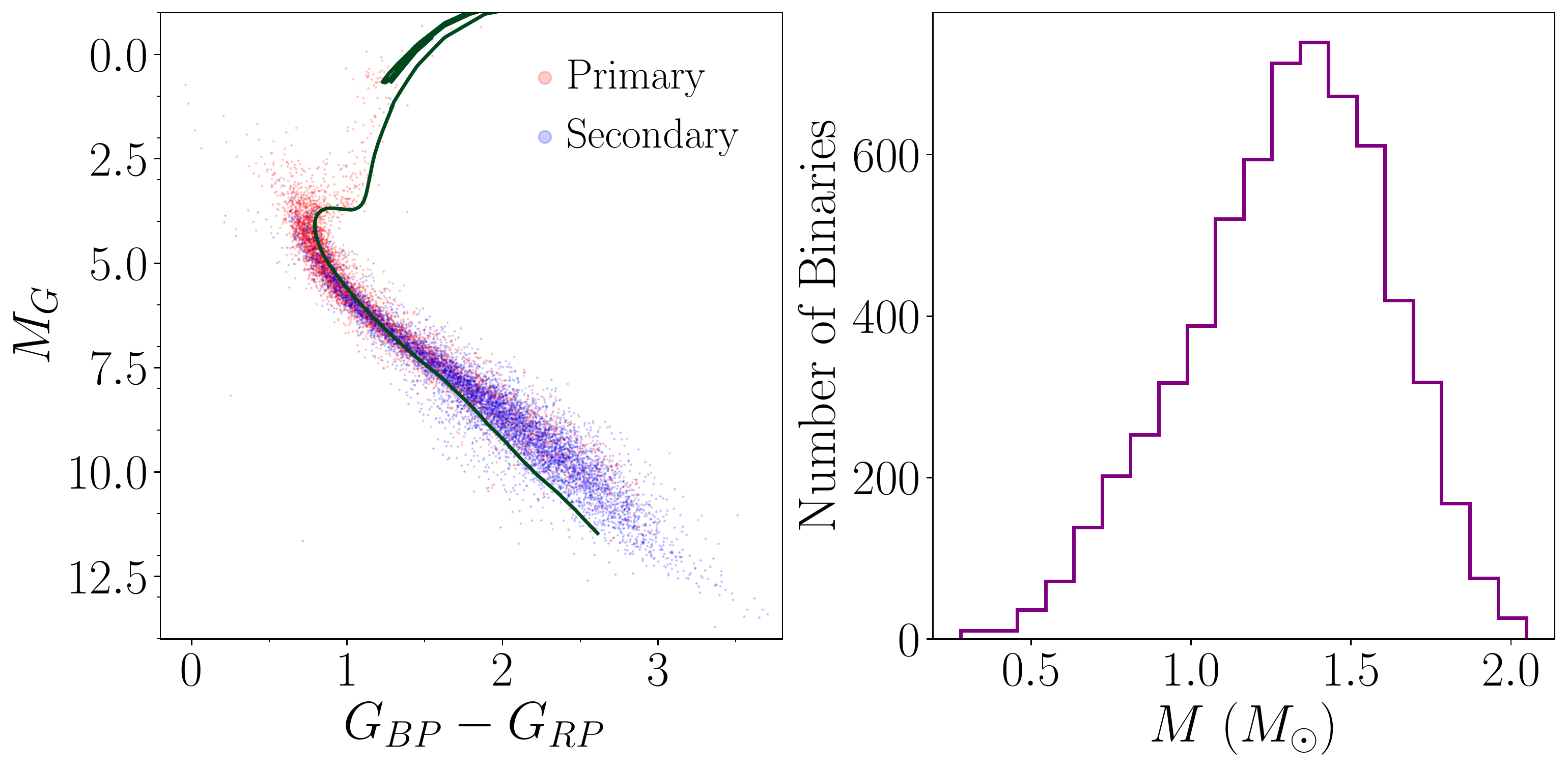}   
    \caption{\textit{(Left)} Color-magnitude diagram of stars in the catalogue corrected for extinction. The stars in a binary that are brighter (fainter) than their stellar companions are labelled as primaries (secondaries). A 10 Gyr \textsc{MIST} isochrone with metallicity ${\rm [Fe/H]} = 0.0$ is shown as a green curve. \textit{(Right)} Binary total mass distribution obtained by interpolating the stellar color-magnitudes over a grid of 10 Gyr \textsc{MIST} isochrones.} \label{fig:mass}
\end{figure*}

As we will describe in Section~\ref{sec:theory}, our limits will in part be set by Monte Carlo simulation of the tidal effects of subhalos on a synthetic population of binaries whose properties match the {\it Gaia} catalogue. 
We calculate the masses of the stellar components (and thus the total mass of each binary system) by considering the sample's extinction-corrected color-magnitude diagram---shown in the left panel of Fig.~\ref{fig:mass}. To correct for extinction, we first calculate the median reddening for each binary using the \textsc{Bayestar 2015} dustmap \cite{bayestar15} implemented within the \textsc{Python} package  \texttt{dustmaps} \cite{dustmaps18}. The reddening values are converted to extinction coefficients corresponding to magnitudes measured in the $G$, $G_{BP}$, and $G_{RP}$ passbands \cite{evans18} using the \texttt{pyia} package \cite{pyia21_1}. Subtracting off these extinction coefficients from their corresponding observed magnitudes gives the intrinsic magnitudes of the stars. To infer the mass of each star from its intrinsic magnitudes, we generate \textsc{MIST} \cite{mist_1,mist_2,mesa_1,mesa_2,mesa_3,mesa_4} isochrones corresponding to stars of age 10 Gyr, for a range of metallicities ${\rm [Fe/H]}$ from $-2.1$ to $+0.4$. For each star, we identify the isochrone closest to the star's position in the color-magnitude diagram. Each of the isochrone's color-magnitude values corresponds to a unique stellar mass. Each star in the sample then is assigned the mass of the closest point within the isochrone.

Since the \textsc{Bayestar 2015} map is only defined for declinations $\delta > -30^{\circ}$, we can only reliably estimate the masses of 6,280 binaries, which form the empirical distribution in the right panel of Fig.~\ref{fig:mass}. 
We use this subset of binaries to approximate the distribution for the total mass corresponding to all 9,637 binaries when we construct our synthetic population.  

\section{Binary Evolution} \label{sec:theory}

Our goal is to place limits on subhalos using the distribution of projected separations of binaries in the catalogue (see Fig.~\ref{fig:data}). Each observed binary was produced with some initial (and unknown) set of orbital parameters, and has evolved over time to its current configuration in part due to the tidal perturbations from subhalos. Given that the initial conditions and history of random tidal encounters are unknown, we must use simulations to determine the statistical distribution of the final orbital parameters of the binaries for a specific population of subhalos. We can then fit simultaneously for the initial orbital distribution of the binaries and the characteristics of the subhalo population.

We first study the effect of a single encounter on the orbit of a binary, followed by the cumulative effect of many random encounters on the binary's orbit. Although the effect of a single encounter on the binary's orbit is deterministic, random encounters only allow us to describe the effect of encounters as a scattering matrix describing the probability that a binary possesses a specific orbit after it encounters a random population of perturbers.

With this scattering matrix, we can then evolve a population of simulated binaries from some initial distribution of projected separations to a final distribution, assuming a set of dark matter perturbers with specified properties. This evolved distribution can be reweighted as the primordial binary distribution is varied, allowing us to set robust limits in Section~\ref{sec:stats} without assuming a particular formation mechanism for the binaries.

\subsection{The Effect of a Single Encounter on a Binary}

A binary star system consists of two stars in a bound Keplerian orbit supported by their mutual gravity. The bound orbit is elliptical and specified by semimajor axis $a$ and eccentricity $e$.  The physical separation $r$ of the stars evolves as
\begin{equation} \label{eq:r}
    r = a(1 - e \cos \psi),
\end{equation}
where $\psi$ is the eccentric anomaly, which is related to the dynamical time $t$ through
\begin{equation} \label{eq:time}
    t = \frac{P}{2 \pi} \left ( \psi - e \sin \psi \right ).
\end{equation}
Here, $P = a^{3/2} \sqrt{4 \pi^{2} / GM}$ is the binary's orbital period and $M$ is the binary's total mass. 

For tidal interactions between the wide binaries of interest and subhalos within the Milky Way, the relative speed of each encounter is high enough that the timescale of the interaction is short compared to the orbital period. Thus, we can invoke the impulse approximation \cite{banik2020, galtext, gnedin99, spitzer58}, which treats $\psi$ and the binary separation $r$ as constant during the interaction. The result is an instantaneous velocity kick that changes a binary's orbit.  

\begin{figure}[t!] 
    \centering
    \includegraphics[width = 0.65\columnwidth]{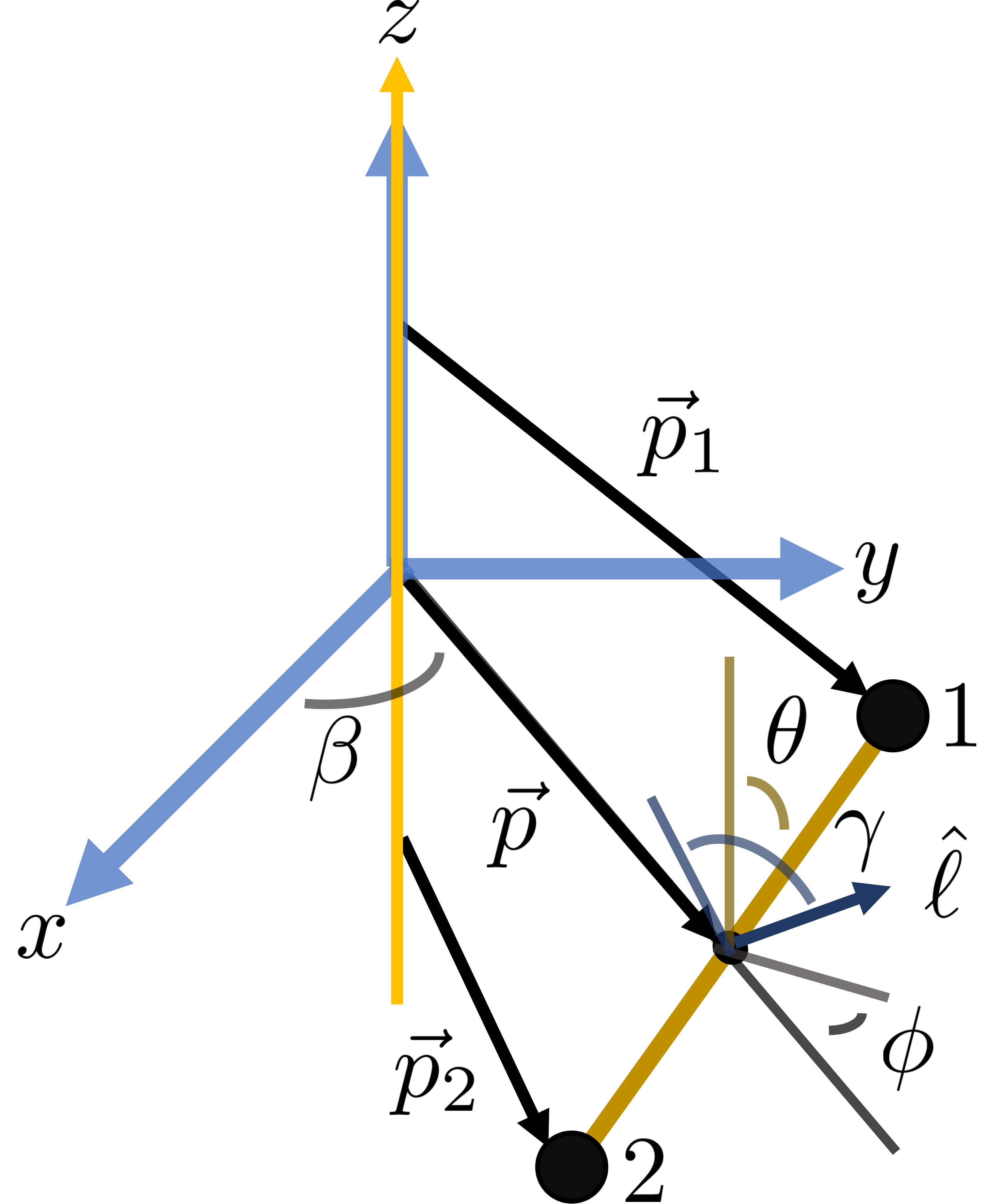}   
    \caption{Interaction geometry of perturber-binary encounters. Each possible encounter is uniquely specified by the position and orientation of the binary relative to the perturber. The former is specified by the impact parameter $p = |\vec{p}|$ and the angle $\beta$. The latter is specified by the angles $(\theta,\phi,\gamma)$, where the $\theta$ and $\phi$ angles describe the orientation of the binary components during the tidal interaction, and $\gamma$ describes the orientation of the binary orbital plane. The binary-perturber interaction is independent of the angle $\beta$. See the text for more details.} \label{fig:pc}
\end{figure}

To calculate the effect of a spherically symmetric subhalo moving past a binary, we use the interaction geometry of Fig.~\ref{fig:pc}. The perturbing subhalo with mass $M_p$ moves with relative speed $v_p$ along the $z$-axis, with an impact parameter $\vec{p}$ to the midpoint of the binary (located in the $x$-$y$ plane). The separation for the stars during the encounter is $r$, which is a function of the orbital parameters $a$ and $e$, as well as the eccentric anomaly $\psi$, all of which are held constant throughout the interaction. During the tidal encounter, the axis connecting the binary components is oriented at angles $\theta$ and $\phi$ relative to $\vec{p}$ and the $z$-axis. Each star therefore has a separate impact parameter $\vec{p}_i$ ($i=1,2$). In addition, an angle $\gamma$ specifies the orientation of the binary's orbital plane relative to the cross product of the binary separation vector and $\vec{p}$. 

\begin{figure}[t!] 
    \centering
    \includegraphics[width = 0.9\columnwidth]{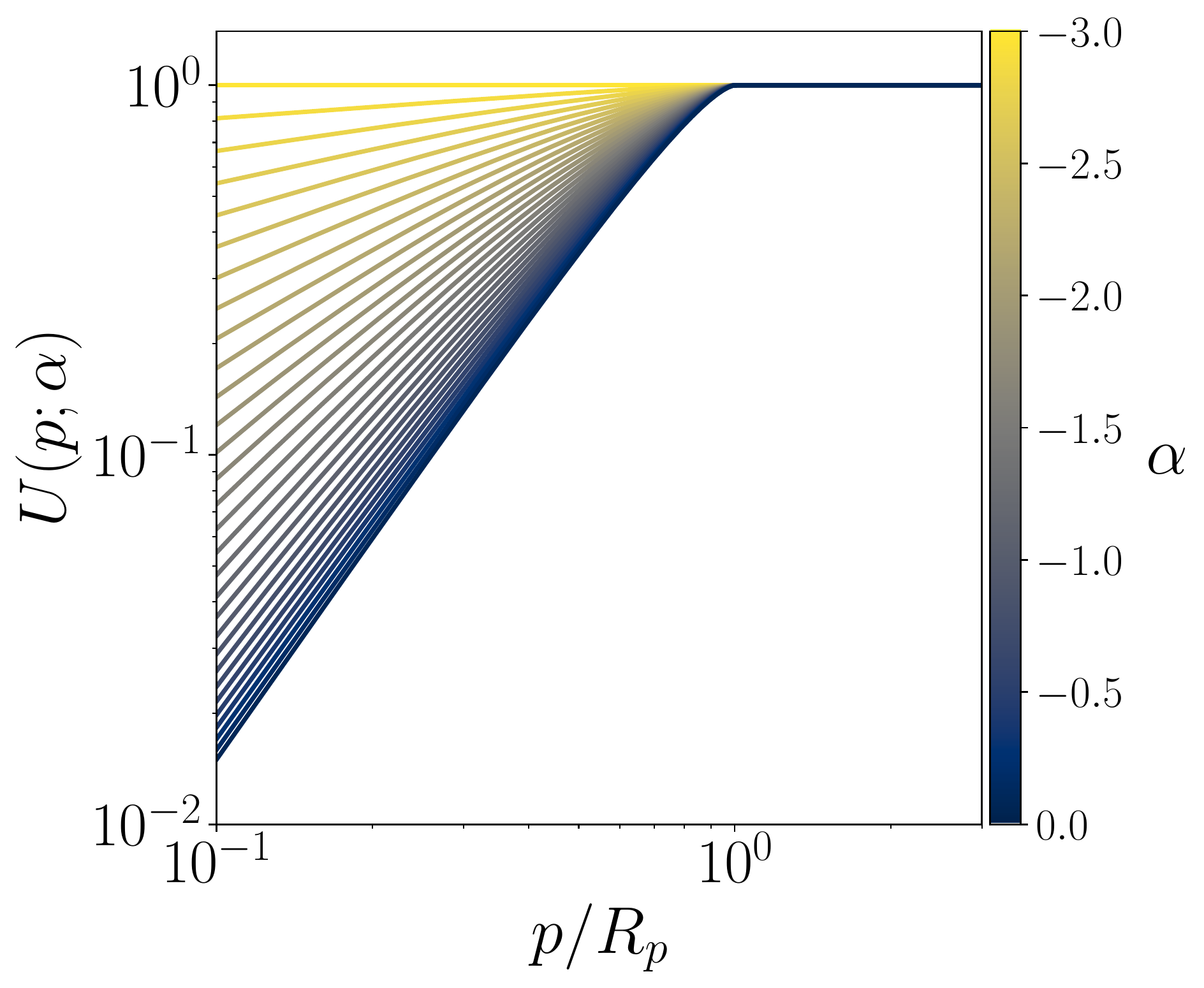}   
    \caption{Structure function for perturbers with different power-law density profiles given by Eq. \eqref{eq:powerlaw_density_function}.} \label{fig:u}
\end{figure}

The velocity kicks imparted on the components of the binary are then \cite{white85}
\begin{equation} \label{eq:impulse}
    \Delta \vec{v}_{i} = -\frac{2GM_{p}}{v_{p}} U(p_{i}) \frac{\vec{p_{i}}}{p_{i}^{2}},
\end{equation}
where the structure function $U(p)$ \cite{gon13,gnedin03} is given by
\begin{equation} \label{eq:u}
    U(p) = \int_1^\infty d\xi \frac{\mu_{p} (p \xi)}{\xi^2 \sqrt{\xi^2-1}}.
\end{equation}
Here, $\mu_{p}(r)$ is the perturber's normalized enclosed mass $M_{p}(<r)/M_{p}$, where $M_p$ is the total mass of the perturber. In Fig.~\ref{fig:u}, we plot the structure functions of perturbers with radius $R_{p}$ and a power-law density profile of the form:
\begin{equation}
    \rho(r;\alpha) = \left\{ \begin{array}{cr} \rho_0 \left(\frac{r}{R_p}\right)^{\alpha}, & r \leq R_p \\ 0, & r>R_p, \end{array}\right. \label{eq:powerlaw_density_function}
\end{equation}
where $\rho_0$ is a characteristic density set by the perturber's total mass $M_p$, and the power-law index is $\alpha > -3$. Note that as $\alpha \to -3$, the $U$ function approaches 1 for all $p$. This is the same structure function as that of a point-mass perturber.

The velocity kicks alter the binary's internal energy per reduced mass $E = -GM/2a$ and internal angular momentum per reduced mass $|\vec{\ell}|  = \sqrt{GM a (1-e^{2})}$. The change in $E$ is
\begin{align} \label{eq:E_impulse}
    \Delta E & = \frac{\Delta v^{2}}{2} + \vec{v} \cdot \Delta \vec{v},
\end{align}
where $\vec{v} = \vec{v}_{1} - \vec{v}_{2}$ and $\Delta \vec{v} = \Delta \vec{v}_{1} - \Delta \vec{v}_{2}$. The change in $\vec{\ell}$ is
\begin{align} \label{eq:l_impulse}
     \Delta \vec{\ell} = \vec{r} \times \Delta \vec{v},
\end{align}
where $\vec{r} = \vec{r}_{1} - \vec{r}_{2}$ is the separation vector of the stars.

\begin{figure}[t!]
    \centering
    \includegraphics[width = 0.9\columnwidth]{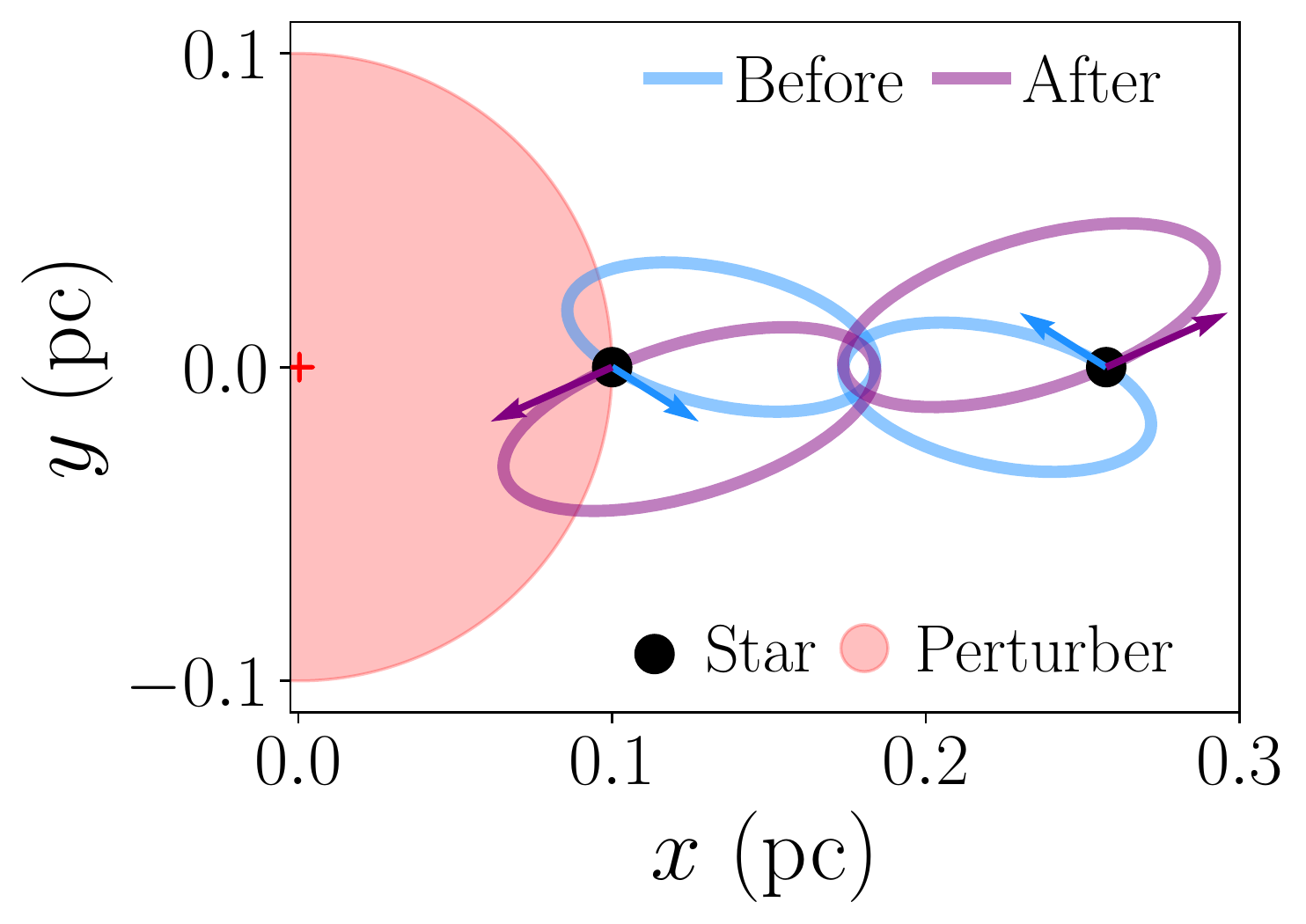}  
    \caption{The result of an encounter between a binary and a perturber. The binary is made up of two $0.5 \ M_{\odot}$ stars orbiting in the plane of the page. The perturber has mass $M_{p} = 10^{3} \ M_{\odot}$, radius $R_{p} = 0.1 \ {\rm pc}$, a uniform density profile, and passes perpendicular to the plane of the page. The state of the binary before the encounter is $(a,e,\psi/2 \pi) = (0.1 \ {\rm pc}, 0.9, 0.64)$. The encounter parameters are $(p, \phi, \theta, \gamma, v_{p}) = (0.17 \ {\rm pc},0,\pi/2,0,240 \ {\rm km/s} )$.} \label{fig:single_encounter}
\end{figure}

If $\Delta E \geq |E|$, the encounter unbinds the binary. If $\Delta E < |E|$, then the binary remains bound, evolving to a new semimajor axis $a' = a+\Delta a$ with 
\begin{equation}
    \frac{\Delta a}{a} = \frac{\Delta E/|E|}{1-\Delta E/|E|},
\end{equation}
and a new eccentricity $e'$ given by 
\begin{equation}
    |\vec{\ell}+\Delta\vec{\ell}| = \sqrt{GM a' (1-e'^{2})}. 
\end{equation}
Up to a minus sign, the new eccentric anomaly $\psi'$ is determined by setting the separation immediately before and after the interaction equal:
\begin{equation}
    a(1-e\cos\psi) = a'(1-e'\cos\psi').
\end{equation}
We determine the sign of $\psi'$ by noting that it shares the same sign as the first time-derivative of the separation $r$. In Fig.~\ref{fig:single_encounter}, we show an example of the change in a single binary's orbit due to the tidal forces from the passage of a single extended perturber of mass $10^3\,M_\odot$, radius $R_p = 0.1$~pc, and constant density.

\subsection{Many Random Encounters on a Single Binary} \label{ssec:multi}

A binary in the Milky Way's stellar halo/thick disk will have encountered many tidal perturbers over its life, each with random orientations, relative velocities, and impact parameters. The frequency of the encounters depends on assumptions about the population of perturbers---the dark matter subhalos in our case---while the ability of any particular interaction to modify the binary's orbit depends on both the perturber population (through their individual masses $M_{p}$ and the structure function $U$), the binary orbital state (semimajor axis $a$, eccentricity $e$, eccentric anomaly $\psi$, and mass $M$), as well as the distance of closest approach and relative orientation of the tidal encounter. If we assume a uniform population of perturbers with an isotropic velocity distribution, the effect of repeated random encounters can be encoded in a scattering matrix \cite{yoo04}. In this subsection, we will develop the scattering matrix formalism, which we will then apply to the population of binaries using Monte Carlo techniques in Section~\ref{ssec:multiple-encounters}.

We denote the state of a binary orbit as $\vec{q} = (a,e,\psi)$. An encounter with a subhalo will alter an initial $\vec{q}_0$ to a new $\vec{q}_1$. For a specified encounter, the evolution from $\vec{q}_0$ is deterministic. However, for the encounter parameters $(p,\theta,\phi,\gamma,v_p)$ of an unspecific random encounter, the final state can only be quantified by the probability distribution of $\vec{q}_1$ given $\vec{q}_0$, $f_1(\vec{q}_1|\vec{q}_0)$. In addition to $\vec{q}_0$, this probability distribution depends on the perturber's structure function and mass, and the distributions from which the encounter parameters are drawn.

\begin{figure*}[t!]
    \centering
    \includegraphics[width = 1.9\columnwidth]{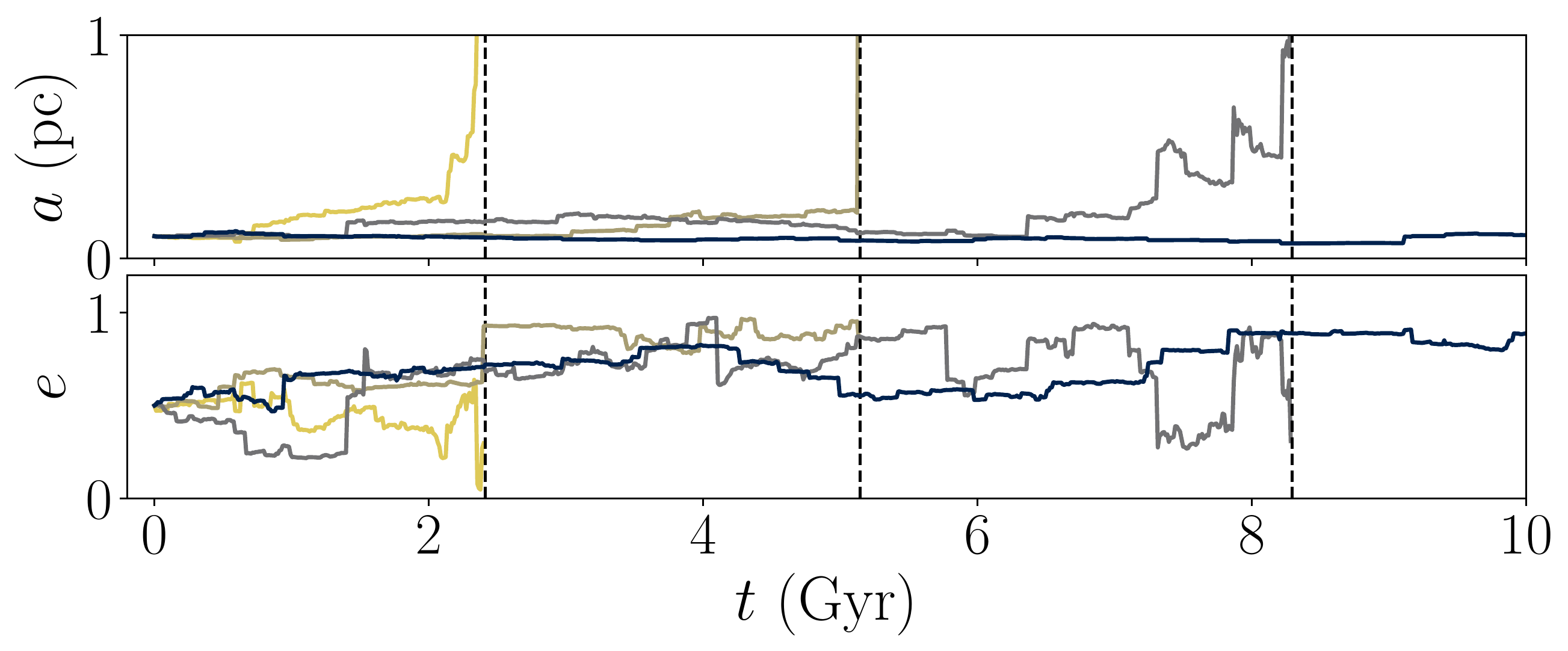}  
    \caption{Semimajor axis and eccentricity evolution (solid lines) of four identical binaries under the influence of uniform-density perturbers with $(M_p,R_{p},f_{p}) = (10^{3} \  M_{\odot}, 0.1 \ {\rm pc}, 1)$. All the binaries have mass $M = 1 \ M_{\odot}$ and are initially in the state $(a_{0},e_{0},\psi_{0}) = (0.1 \ {\rm pc}, 0.5, 0)$. They evolve for $10 \ {\rm Gyr}$. Disruption times are denoted by the dashed vertical lines.} \label{fig:single_binary}
\end{figure*}

For repeated encounters, the probability distribution $S_2$ of the orbital state after the second encounter, $\vec{q}_{2}$, is the probability $f_{1}$ of $\vec{q}_0$ evolving to some intermediate $\vec{q}_1$ followed by the probability $f_{2}$ of evolution of $\vec{q}_1$ into $\vec{q}_2$ (including the change in the eccentric anomaly $\psi$ due to orbital evolution between the first and second encounter), integrated over all possible intermediate states:
\begin{equation}
    S_2(\vec{q}_2|\vec{q}_0) = \int d\vec{q}_1 \ f_{2}(\vec{q}_2|\vec{q}_1)f_{1}(\vec{q}_1|\vec{q}_0).
\end{equation}
This can be continued for an arbitrary number of encounters, resulting in the scattering matrix $S$, which gives the probability distribution of the final orbital state $\vec{q} \equiv \vec{q}_{N}$ given the initial $\vec{q}_0$ and $N$ encounters between the binary and subhalos over a total time $T$:
\begin{equation}
    S(\vec{q}|\vec{q}_0) = \int\prod_{i=1}^{N-1}\Big[ \ d\vec{q}_i \ f_{i+1}(\vec{q}_{i+1} |\vec{q}_i)  \Big] \ f_{1}(\vec{q}_{1} |\vec{q}_{0}),
\end{equation}
where the product of integrals appears from there being $N-1$ intermediate states.

In practice, we calculate the scattering matrix via Monte Carlo simulations. For a binary with initial parameters $\vec{q}$, we draw $N$ random encounters over a total time $T$ assuming uniform spacing between encounters $\delta t \sim T/N$, allowing the binaries to evolve along their new orbits after each encounter.\footnote{A more physically motivated assumption would be to simulate encounters with a random time-step rather than uniform. However, this choice would limit our ability to parallelize our code. We have compared results using both random and uniform spacing between encounters and found them to be identical.} We set $T = 10 \ \rm{Gyr}$, consistent with the age of the stellar halo/thick disk \cite{carroll96}, so that scattering occurs from the time that the entire binary population was assembled to the present-day.

The subhalo population is assumed to be homogeneous, all with the same mass $M_p$ and a specified density distribution (we will consider various possible density profiles in Section~\ref{sec:results}). In this section, we take perturbers with mass $M_{p} = 10^{3} \ M_{\odot}$, radius $R_{p} = 0.1 \ {\rm pc}$, and uniform density profile ($\rho(r) = {\rm constant}$) as our working example. We assume that both the binaries and the perturbing subhalos are moving in the stellar halo\footnote{As the disk stars are rotationally-supported, thick disk binaries are more likely to experience lower-velocity encounters than halo binaries. Since lower-velocity encounters lead to stronger velocity kicks, assuming thick disk binaries have the same velocity distribution as halo binaries leads to conservative limits.} with isotropic Maxwell-Boltzmann velocity distributions---each with velocity dispersion $\sigma$ satisfying $\sqrt{2}\sigma = 200$~km/s \cite{yoo04,galtext} and truncated at the local escape velocity, $v_{\rm esc} = 533$~km/s \cite{piffl2014,zyla2020}. Similar to Ref.~\cite{yoo04}, these distributions yield a relative velocity distribution consistent with subhalos moving in an isothermal sphere with circular velocity $v_{c} = 220 \ {\rm km/s}$ and halo binaries with velocity dispersions given by $(\sigma_\pi, \sigma_\theta, \sigma_Z) = (153,106, 101)$ km/s, as measured in RR Lyrae stars moving in the stellar halo \cite{muhie_2021}. We specify the number density of the subhalos as the fraction $f_{p}$ of the total local dark matter density they compose, assuming $\rho_{\rm DM} = 0.0104\,M_\odot/{\rm pc}^3$ \cite{catena_2010,read2014,zyla2020}. Thus, the local number density is specified by $f_{p}$ and $M_p$. Under these assumptions, the impulse approximation for tidal encounters is accurate up to subhalo masses of $\sim 10^8\,M_\odot$ \cite{yoo04}.

During the Monte Carlo simulations, we draw random encounter parameters corresponding to the impact parameter, relative velocity, orientation of the binary separation relative to the subhalo trajectory, and relative orientation of the binary's orbital plane for each encounter. The velocities $v_{p}$ are sampled from the assumed Maxwell-Boltzmann distributions, the separation orientation angles $(\theta,\phi)$ are sampled uniformly from the solid angles $\Omega$, the orbital plane orientation angles $\gamma$ are sampled uniformly from $0$ to $2\pi$, while the impact parameters $p$ are sampled uniformly from the disk around the binary's midpoint. The maximum impact parameter $p_{\rm max}$ sampled is defined as the impact parameter at which the expected cumulative set of tidal interactions---each with impact parameters $>p_{\rm max}$---between the subhalo and the binary can, at maximum, inject 1\% of the binary's initial binding energy over time $T$ (assuming circular binary orbits and perturber velocity perpendicular to the binary).  We have verified that our results are robust to this choice of $p_{\rm max}$. 

To calculate the expected number of tidal encounters $N$, we first note that the scattering rate $d\mathcal{N}/dt$ of a binary interacting with perturbers with fixed relative velocity $v_{p}$ depends on the subhalo mass $M_{p}$, fraction of the dark matter density composed of perturbers $f_{p}$, and $p_{\rm max}$ as
\begin{equation}
     \frac{d \mathcal{N} }{dt} = f_{p} \left ( \frac{\rho_{\rm DM}}{M_{p}} \right ) \times \pi p_{{\rm max}}^{2} \times v_{p}.
\end{equation}
The expected time between encounters is $\delta t$ = $(d \mathcal{N}/dt)^{-1}$. Given our relative velocity distribution, we calculate the velocity-averaged time between encounters, $\langle \delta t \rangle$, from which we can calculate the expected number of encounters in time $T$ as
\begin{equation}
    N = {\rm int} \left[ \frac{T}{\left \langle \delta t \right \rangle} \right] . \label{eq:NfT}
\end{equation}

In Fig.~\ref{fig:single_binary}, we show the evolution of four example $M = 1 \ M_{\odot}$ binaries over 10~Gyr. All four began in the same initial state $(a_{0},e_{0},\psi_{0}) = (0.1 \ {\rm pc},0.5,0)$ and interacted with a population of uniform-density subhalos with $(M_p,R_{p},f_{p}) = (10^{3} \ M_{\odot}, 0.1 \ {\rm pc}, 1)$. As the four binaries randomly interact with perturbers, their orbits evolve in different ways. While individual binary-subhalo interactions can increase or decrease the semimajor axis, the general trend can be seen to be one of gradually widening binaries from tidal heating. In this particular set of examples, three out of four of the binaries end with complete disruption ($a \to \infty$) as a final encounter leads to energy injection above the binding energy, while the fourth case experiences only negligible changes in its semimajor axis.

\begin{figure}[t!]
    \centering
    \includegraphics[width = 0.9\columnwidth]{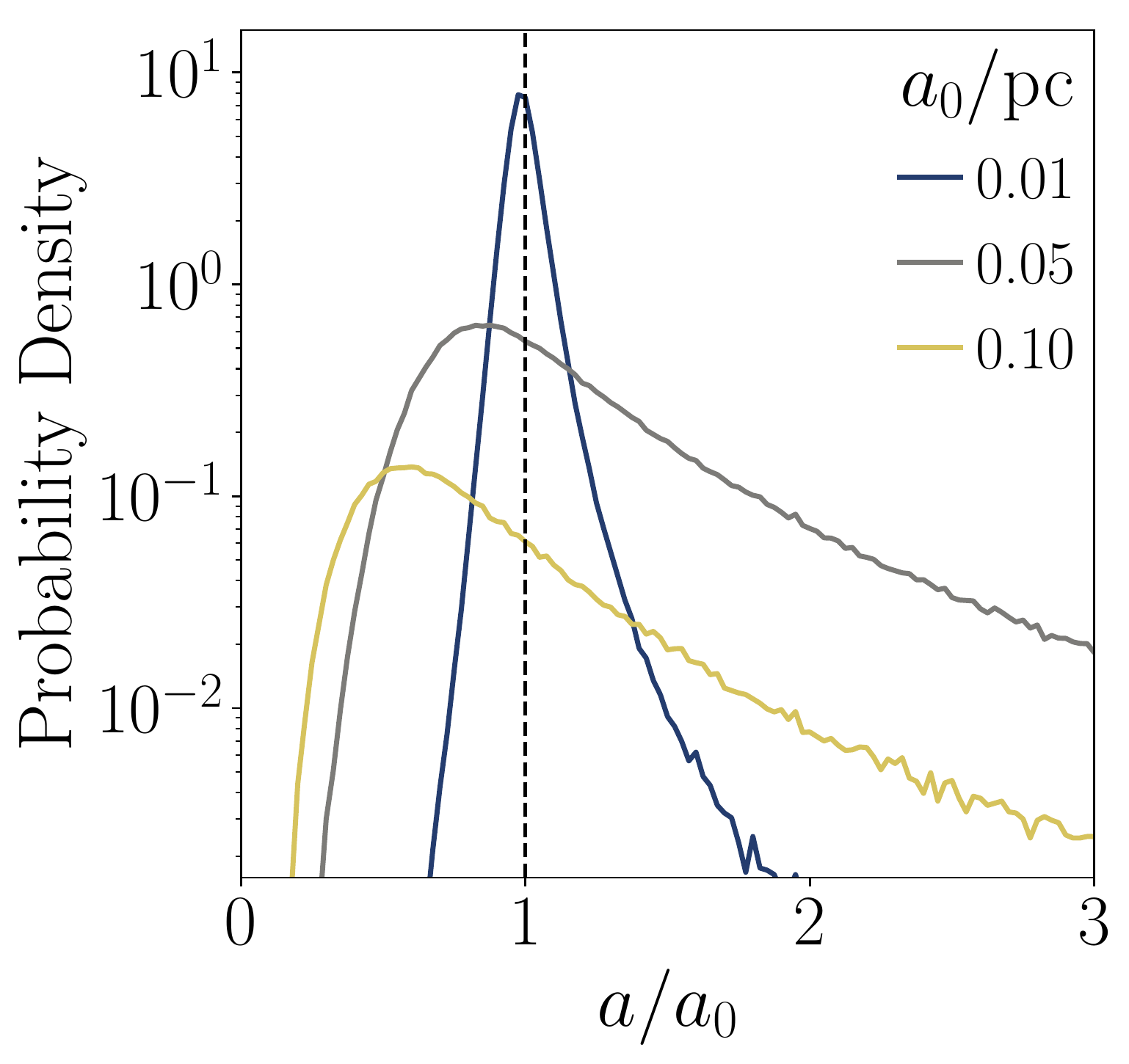}  
    \caption{Probability density of scattering to semimajor axis $a$ assuming $e_0 = 0.5$, $\psi_0/2\pi = 0$, and $M = 1 \ M_{\odot}$ after 10 Gyr evolution with a population of uniform-density perturbers with $(M_p,R_{p},f_{p}) = (10^{3} \  M_{\odot}, 0.1 \ {\rm pc}, 1)$ for three initial semimajor axes: $a_0 = 0.01$, $0.05$, and $0.1$~pc. Each distribution was obtained using $10^{6}$ simulated binaries.} \label{fig:single_a}
\end{figure}

In Fig.~\ref{fig:single_a}, we show the final distribution of $a$ given initial $e_{0}=0.5$, $\psi_{0} = 0$, and initial semimajor axes of $a_{0}=0.01$, $0.05$, and $0.1$~pc for binaries with $M = 1 \ M_{\odot}$. Each distribution was calculated using $10^{6}$ Monte Carlo simulations. These are the scattering matrices for the semimajor axis $a$ corresponding to the three initial binary configurations, marginalized over the rest of the orbital elements. From these results, we see that, as the semimajor axis of a binary increases, interactions with the subhalo population can more easily increase $a$.

\subsection{The Effect of Many Random Encounters on Multiple Binaries} \label{ssec:multiple-encounters}

The scattering matrix formalism---which describes the probability distribution of the binary's final orbit $(a,e,\psi)$ given a specified initial orbit and a population of subhalo perturbers---can now be applied to a population of binaries that themselves have a range of initial conditions. Our ultimate goal is to compare a predicted distribution with measurable parameters within the {\it Gaia} wide binary catalogue; to that end, we will construct a probability distribution of the observed separation $s$ between the stellar components of the binaries.

We denote the initial probability distribution of the orbital state $\vec{q}_0$ as $\phi_0(\vec{q}_{0})$. As our sample consists of binaries with different masses, we redefine the binary state to include the binary mass $M$: $\vec{q} \rightarrow (a,e,\psi,M)$. Unlike the other elements of the state vector, $M$, while affecting the evolution of a binary, does not change during the evolution. With this change in notation, the present-day probability distribution of $\vec{q}$ after experiencing encounters with a population of subhalos over time $T$, is 
\begin{equation}
    \phi(\vec{q}) = \int d\vec{q}_0 \ S(\vec{q}|\vec{q}_0) \ \phi_0(\vec{q}_0). \label{eq:sdist}
\end{equation}

To calculate the present-day distribution $\phi$, we must specify initial distributions for the binary orbital state parameters. The initial distribution of the semimajor axes of wide binaries is not well understood, though it is generally taken to be a power law \cite{wasserman87,weinberg87,opik24}. As a result, we will not specify this distribution {\it a priori}. Rather, we assume it obeys a power law and marginalize our constraints over the power-law index---in Appendix~\ref{app:sbplaw}, we consider the possibility that the initial semimajor axis distribution is a broken power law, as in T19. 

We then calculate $\phi(\vec{q})$ over narrow ranges of $a_0$ (assuming uniform distributions within this range). In Section~\ref{sec:stats}, when we place observational limits on a population of subhalos given our sample of binaries, we can then vary the initial distribution of the semimajor axes by reweighting each range of $a_0$. 

The initial distribution of eccentricities $e_0$ is usually taken to be either thermal, $\phi_{0}(e_0) = 2e_0$, or superthermal, $\phi(e_0) \propto e_0^{\kappa}$ (where $\kappa > 1$) \cite{weinberg87,jeans1919,geller19,hwang2021}. The {\it Gaia} wide binaries from E21 have a present-day distribution of eccentricities that is consistent with the superthermal exponent $\kappa$ increasing from $\kappa=1$ as the semimajor axis increases \cite{hwang2021}, though the full behavior of this distribution is not well-characterized.
Highly-eccentric orbits will be more susceptible to disruption during tidal encounters with a subhalo (due to the greater amount of time binaries spend around their apocentric phases). Such orbits are more common in superthermal distributions, and so, to place conservative limits, we adopt the thermal distribution for our initial eccentricities. 

The eccentric anomaly $\psi$ in the {\it Gaia} catalogue of wide binaries is not directly observable. However, the initial phases of the binaries $\psi_{0}$ are expected to be randomly distributed in dynamical time $t$ with uniform probability. Therefore, from Eq.~\eqref{eq:time}, the conditional probability of $\psi_0$ given $e_0$ is \begin{equation}
    \phi_{0}(\psi_0| e_0) = \frac{1}{2 \pi} (1 - e_0 \cos \psi_0).
\end{equation} 
The initial distribution of masses $M$ is given by the empirical mass distribution to the right of Fig. \ref{fig:mass}.

\begin{figure*}[t!] 
    \centering
    \includegraphics[width = 1.8\columnwidth]{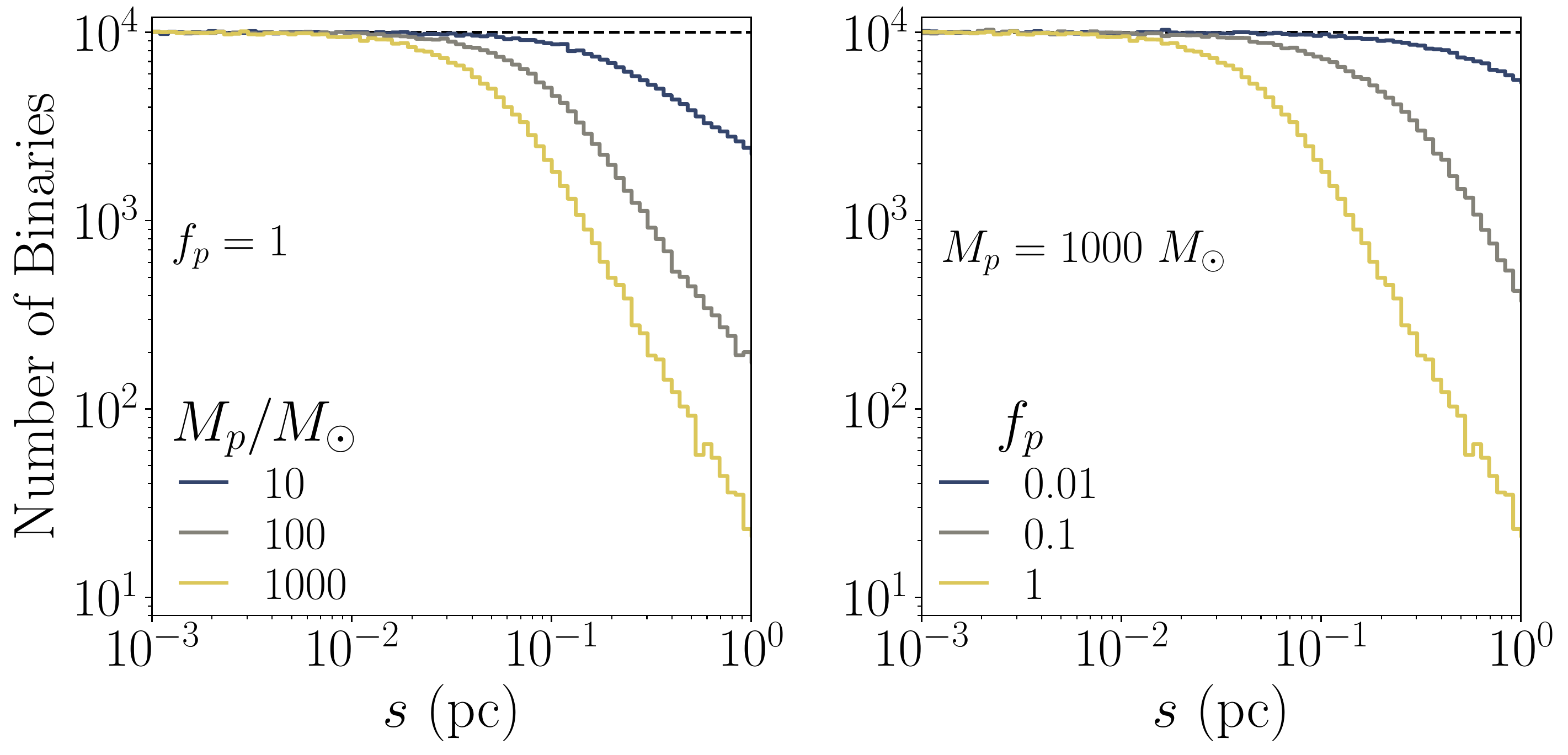}   
    \caption{Number of binaries per bin of logarithmic projected separation for binary populations that each have evolved with a different set of 0.1 pc uniform-density perturbers for 10 Gyr (solid lines) and were initially distributed uniformly in logarithmic semimajor axis (dashed line). \textit{(Left)} Perturbers have various masses $M_{p}$ and $f_{p} = 1$. \textit{(Right)} Perturbers have mass $M_{p}= 10^{3} \ M_{\odot}$ and various perturber fractions $f_p$.} \label{fig:flat}
\end{figure*}

The most directly measurable property of the wide binaries in the {\it Gaia} catalogue is not the semimajor axis, eccentricity, or eccentric anomaly. Rather, it is the projected separation $s$ of the binaries at the time of observation. It is related to the physical separation $r$ through the line-of-sight inclination angle of the binary $i$:
\begin{equation}
    s = r\cos i,
\end{equation}
where $r$ is related to the orbital state $\vec{q}$ through Eq.~\eqref{eq:r}. We assume binaries are uniformly distributed in $\sin i$, as the orientation of the binaries is uncorrelated with their line-of-sight to Earth \cite{wasserman87}. The probability distribution for $s$ is then
\begin{equation}
    \phi(s) = \int d\sin i \int d\vec{q} \ \delta(s-r\cos i) \ \phi(\vec{q}),
\end{equation}
where $\delta$ denotes the Dirac delta function.

As an example, we show in Fig.~\ref{fig:flat} the numerically-derived distributions for $\phi(s)$ assuming an initial distribution of semimajor axes which is uniform in log-space. For our example subhalo population, we continue using uniform-density subhalos with radius $R_{p} = 0.1~{\rm pc}$. We repeat the numerical calculation for different choices of perturber mass $M_{p}$ and perturber fraction $f_{p}$. For these numeric calculations, we generate binaries with semimajor axes sampled uniformly across 175 bins logarithmically spaced between $a_{0} = 10^{-5}$~pc and $a_{0} = 10^2$~pc.\footnote{This range is larger than the $10^{-4}-10^0$~pc range of the wide binary catalogue, to allow for binaries migrating into the region of interest as a result of tidal encounters.} Each bin contains $10^{4}$ binaries. After evolving the binaries with subhalos for 10~Gyr, the initially flat distribution in $s$ develops a characteristic break at large separations, due to the energy injection from the perturbers. It is this deficit of the widest binaries that will allow us to set limits on the dark matter substructure in Section~\ref{sec:stats}.

\begin{figure}[t!] 
    \centering
    \includegraphics[width = 0.9\columnwidth]{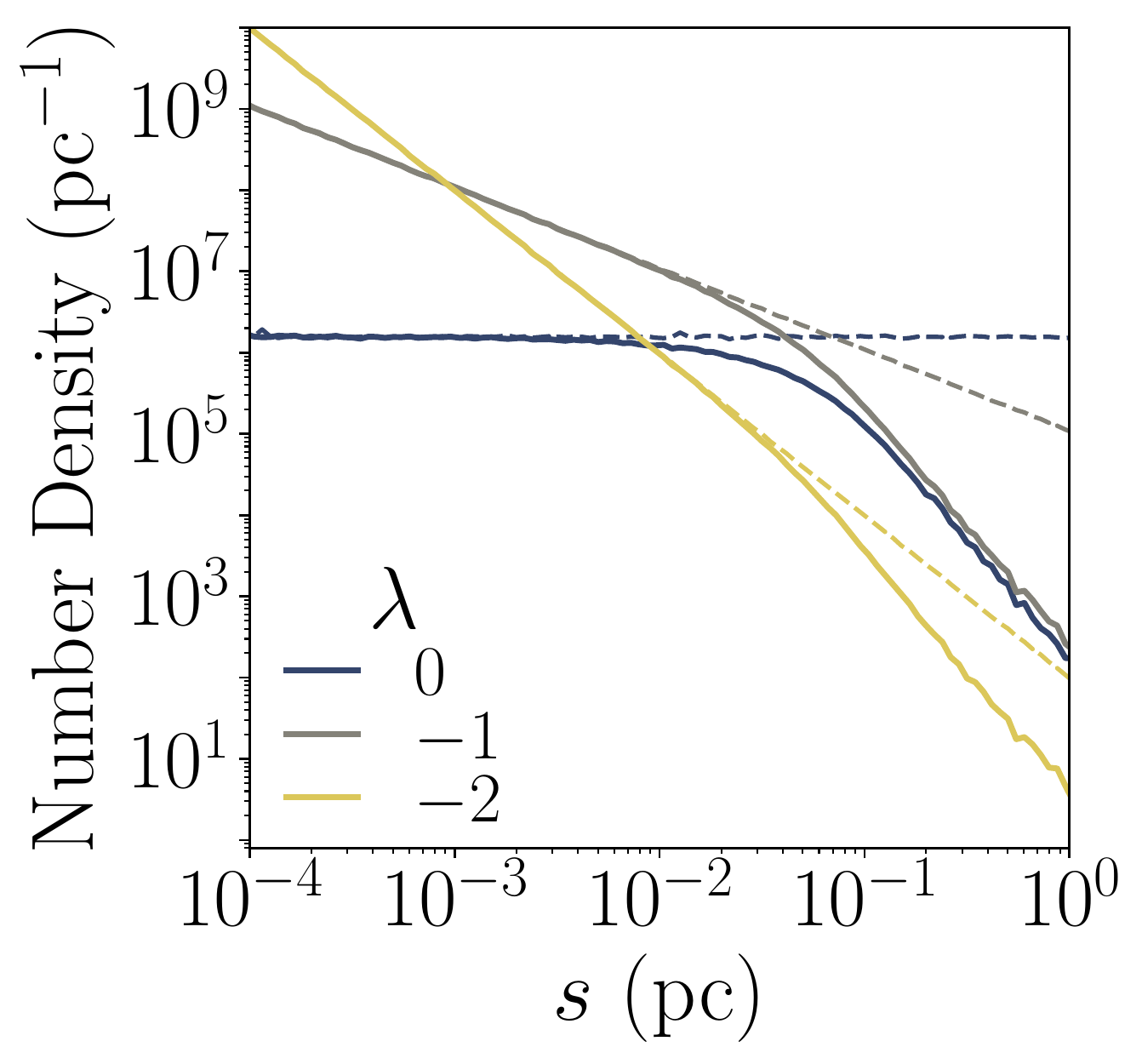}   
    \caption{\textit{Dashed Lines:} Initial projected separation distribution of three populations of binaries with different initial semimajor axis distributions, each obeying different power laws and normalized to $10^{6}$ binaries. \textit{Solid Lines:} Projected separation distributions of the various populations of binaries after they experience encounters with uniform-density perturbers with $(M_{p}, R_{p}, f_{p}) = (10^{3} \ M_{\odot}, 0.1 {\rm \ pc}, 1)$ for $10 \ \rm{Gyr}$.} \label{fig:powerlaws}
\end{figure}

Though the distribution of $s$ has been numerically calculated from samples drawn from a flat distribution of $a_{0}$ in log-space ($\phi_{0}(a_{0}) \propto a_{0}^{-1}$), the behavior of $\phi(s)$ under different assumptions of $\phi_0(a_{0})$ can be straightforwardly calculated by reweighting the binaries based on their initial semimajor axis using Eq.~\eqref{eq:sdist}. In Fig.~\ref{fig:powerlaws}, we show the initial and final distribution of $s$ for three different power-law distributions of initial semimajor axis: $\phi_0(a_{0} | \lambda) \propto a_{0}^{\lambda}$ for $\lambda = 0$, $-1$, and $-2$. These results indicate that the asymptotic behavior of the power law past the break induced by the perturbers is independent of the initial semimajor axis distribution. For the remainder of this work, we will assume the initial probability distribution for the semimajor axis is drawn from a power law with index $\lambda$, with the value of $\lambda$ fit to data, as we will describe in the next section. 

\section{Statistical Methods} \label{sec:stats}

In the previous section, we determined how binary orbits evolve when they are subject to random encounters with subhalos and numerically calculated a scattering matrix that can be integrated over the initial distribution of binaries to give the present-day probability distribution of binary projected separations. We must next compare our calculation of the predicted separation distribution with the observed separation distribution of our sample binaries in order to set limits on the population of subhalos. In this section, we will demonstrate our approach using a single type of subhalo population with uniform density distributions. We will consider other models of dark matter perturbers in Section~\ref{sec:results}.

Previously, we have calculated the probability distribution for the binary projected separation $s$, given tidal interactions over time $T$ originating from a population of subhalos composing a fraction $f_{p}$ of the dark matter density and the power law of the initial semimajor axis distribution $\lambda$. To make these dependencies explicit, we write the present-day distribution as $\phi(s) \rightarrow \phi(s|\lambda, f_{p},\vec{\zeta})$, where we have introduced a parameter vector $\vec{\zeta}$ encoding all other information about the population of subhalos, e.g., the perturbers' masses, radii, and density profiles. 

As stated previously, we wish to set limits on the subhalo abundance $f_{p}$ marginalized over the possible semimajor axis distributions $\lambda$ while keeping the other perturber properties $\vec{\zeta}$ fixed. However, the power-law distribution does not account for \textit{Gaia}'s sensitivity to binaries at different separations or the selection criteria we made in Section~\ref{sec:catalog}. For our sample to be complete at low separations, we required $f_{\Delta G} > 0.999$. This amounts to setting an angular separation cutoff $\theta_{\Delta G}$, depending on the difference in the binary component magnitudes $\Delta G$. Including this selection effect, the probability of detecting and selecting a binary located a distance $d$ from Earth with projected separation $s$ is \cite{ebr2018}
\begin{align}
    p_b(s | d, \Delta G ; \lambda, f_{p}, \vec{\zeta}) = \frac{\phi(s |\lambda, f_{p}, \vec{\zeta}) \ \Theta(s/d - \theta_{\Delta G})}{\int ds' \ \phi(s'|\lambda, f_{p}, \vec{\zeta})  \ \Theta(s'/d - \theta_{\Delta G})},
\end{align}
where $\Theta$ is the Heaviside theta function. 

Moreover, as discussed in Section~\ref{sec:catalog}, not every pair of stars in the binary catalogue is truly a binary. To account for the presence of chance alignments in our sample, we model their separation distribution with a power law, $\phi_{c}(s | \lambda_{c}) \propto s^{\lambda_{c}}$ (we consider other fitting functions in Appendix~\ref{app:chance-align}), and subject them to the same selection effects as the binaries. The probability of detecting and selecting a chance alignment located a distance $d$ from Earth with projected separation $s$ is 
\begin{align}
    p_{c}(s | d, \Delta G ; \lambda_{c}) = \frac{\phi_{c}(s | \lambda_{c}) \ \Theta(s/d - \theta_{\Delta G})}{\int ds'  \ \phi_{c}(s'| \lambda_{c})  \ \Theta(s'/d - \theta_{\Delta G})}.
\end{align}
Using the above two distributions, the probability of having either a binary or a chance alignment in our catalogue is
\begin{align} \label{eq:probability_final}
     p(s|d,& \Delta G, \mathcal{R}; \lambda, \lambda_{c}, f_{p}, \vec{\zeta}) = \\ 
    & (1 - \mathcal{R}) \ p_b(s | d, \Delta G ;  \lambda, f_{p}, \vec{\zeta}) +  \mathcal{R} \ p_{c}(s | d, \Delta G ; \lambda_{c}),  \nonumber
\end{align}
where $\mathcal{R}$ denotes the probability that a selected pair of stars is a chance alignment. As suggested by our notation, for this we use the contamination probability estimate discussed in Section~\ref{sec:catalog}. 

\begin{figure}[t!] 
    \centering
    \includegraphics[width = 0.9\columnwidth]{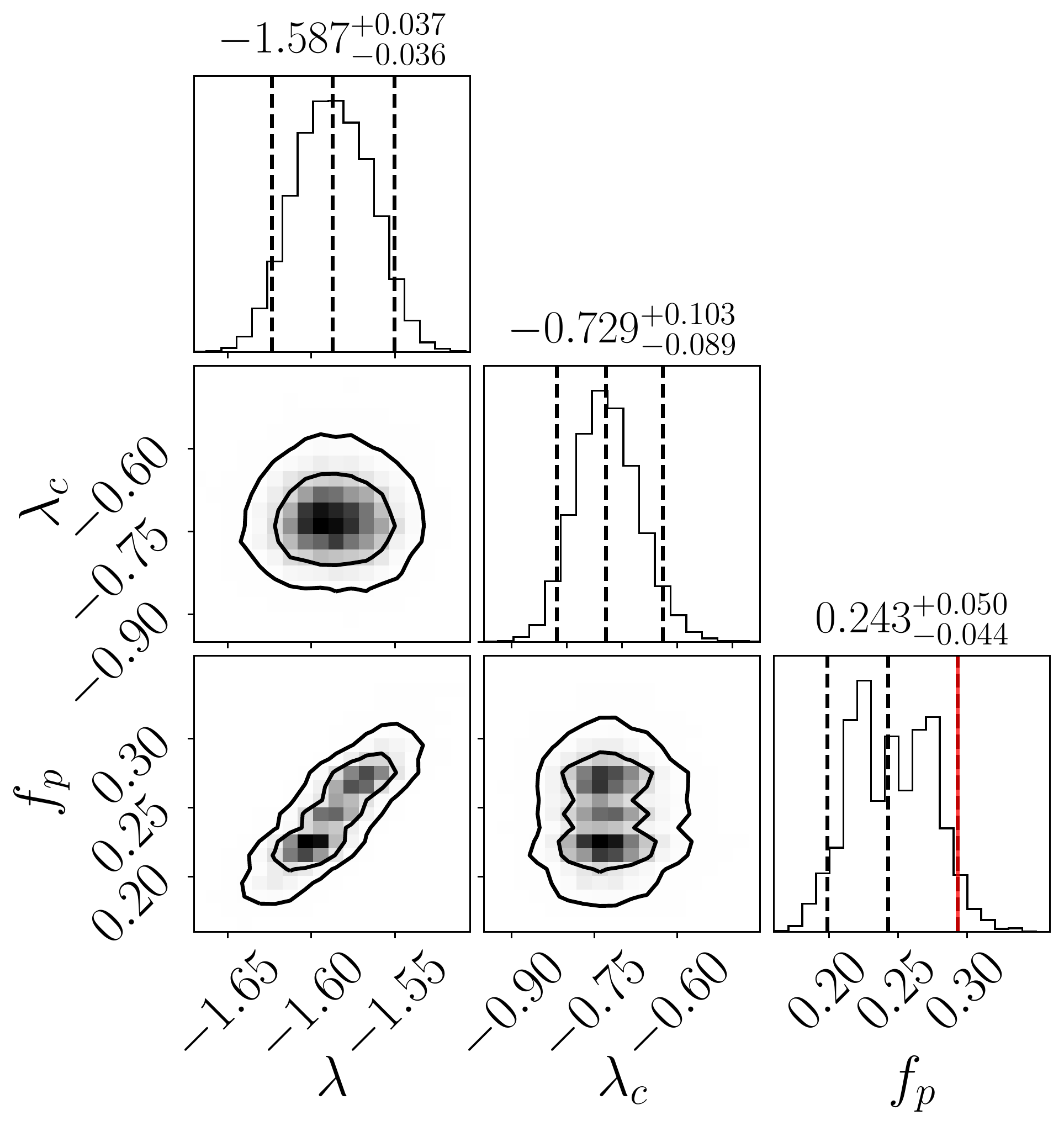}  
    \caption{Sampled posterior distribution of the model parameters $\{ \lambda, \lambda_{c}, f_{p} \}$ for uniform-density perturbers with $(M_{p}, R_{p}) = (10^{3} \ M_{\odot}, 0.1 \ {\rm pc})$. The vertical dashed lines in the 1D histograms denote 5\%, 50\%, and 95\% quantiles. The vertical red line corresponds to our limit on the perturber fraction. The inner and outer boundaries of the 2D contours denote 68\% and 95\% error contours.} \label{fig:pm}
\end{figure}

With the probability distribution given by Eq.~\eqref{eq:probability_final}, we can now calculate a likelihood function ${\cal L}$ of our {\it Gaia} eDR3 wide binary sample, as a function of the perturber fraction $f_{p}$ corresponding to a subhalo population described by the parameters $\vec{\zeta}$, the initial semimajor axis distribution $\phi_0(a_0|\lambda) \propto a_{0}^{\lambda}$, and the population of chance alignments, $\phi_{c}(s | \lambda_{c})\propto s^{\lambda_{c}}$. Assuming the binaries do not affect each other's evolution or detectability, the likelihood function is
\begin{equation}
    {\cal L} = \prod_{i} p(s_{i} | d_{i}, \Delta G_{i}, \mathcal{R}_{i}; \lambda, \lambda_{c}, f_{p}, \vec{\zeta}),
\end{equation} 
where the index $i$ labels the binaries within the sample.

From this, we use Bayes' Theorem to infer the posterior distribution for the model parameters $ \{ \lambda, \lambda_{c}, f_{p} \}$, given the data $\{ s_{i}, d_{i}, \Delta G_{i}, \mathcal{R}_{i} \}_{i}$. We set a limit on the fraction $f_{p}$ of the dark matter composed of subhalos specified by the fixed set of parameters $\vec{\zeta}$. In practice, we sample the posterior distribution using the \texttt{emcee} code \cite{mackey12}, assuming uniform priors for $ \{ \lambda, \lambda_{c}, \log f_{p} \}$, and marginalize over the power-law indices $\lambda$ and $\lambda_{c}$ to obtain the probability distribution for the perturber fraction $f_{p}$. In this way, we report our limit as a 95\% probability bound of the perturber fraction $f_{p}$.

\begin{figure}[t!] 
    \centering
    \includegraphics[width = 0.9\columnwidth]{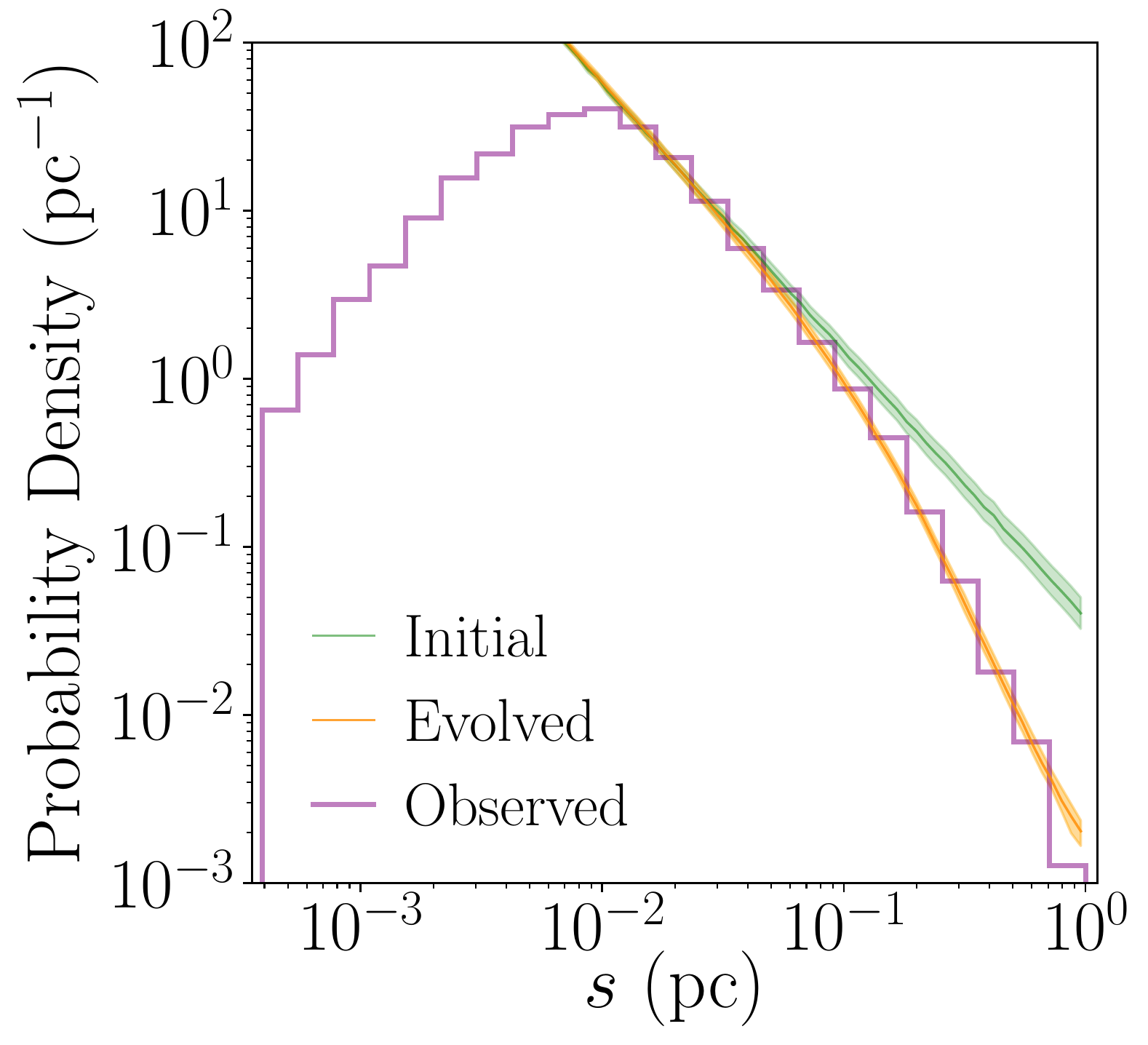}   
    \caption{Fit of the model binary population from Fig.~\ref{fig:pm} to the sample binaries. The expected probability density of observed binaries is given as a histogram produced by weighting each binary candidate with the probability that it is a true binary: $1 - \mathcal{R}$. The best-fit initial and evolved separation distributions are denoted as solid lines. The bands around those lines denote $95\%$ uncertainties around the best-fit model parameters.} \label{fig:pm-fit}
\end{figure}

A sample of the posterior distribution corresponding to a population of uniform-density subhalos with mass $M_{p} = 10^{3} \ M_{\odot}$ and radius $R_{p} =  0.1 {\rm \ pc}$ is shown in Fig.~\ref{fig:pm}. We find the perturber fraction is constrained by the data to be $f_{p} < 0.28$ at the 95\% level, indicated by the solid vertical line at the right end of the distribution for $f_{p}$. In Fig.~\ref{fig:pm-fit}, we show the 
initial power-law distribution of binary separation as well as the evolved final distribution, overlaid on the data. The deviation at low separations is mainly due to the selection cut $f_{\Delta G} > 0.999$. 
We note that our best fit for the unbroken power-law index $\lambda$ is consistent with the results of T19, and the chance-alignment power-law index $\lambda_{c}$ is roughly independent of the perturber population. 

\section{Results} \label{sec:results}

\begin{figure}[t!] 
    \centering
    \includegraphics[width = 0.9\columnwidth]{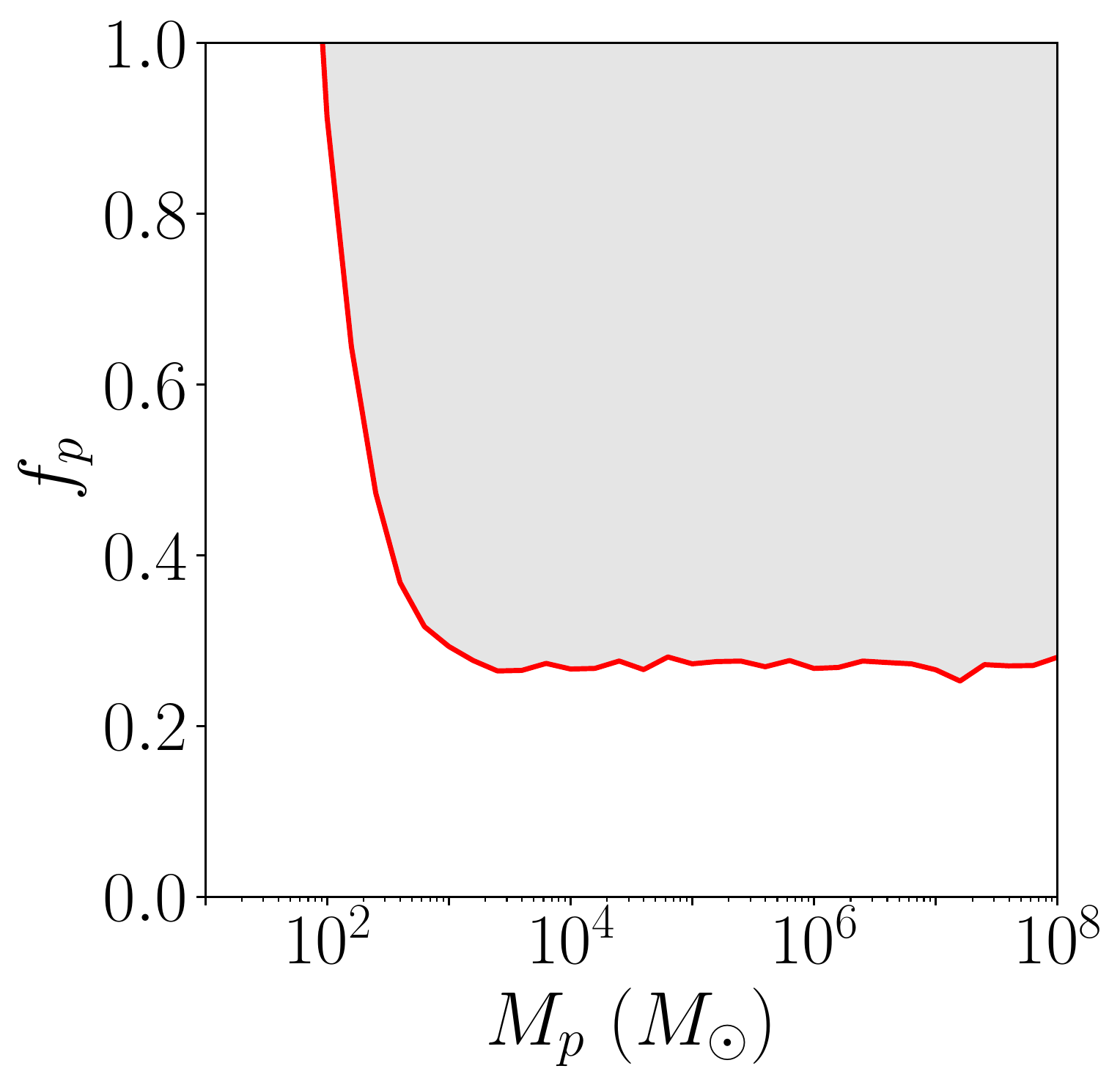}   
    \caption{Limits on populations of uniform-density perturbers with different masses $M_{p}$ and $R_{p} = 0.1 \ \rm{pc}$. The line and shaded area denote the 95\%-excluded region.} \label{fig:constraints_vs_Mp}
\end{figure}

\begin{figure*}[t!] 
    \centering
    \includegraphics[width = 1.95\columnwidth]{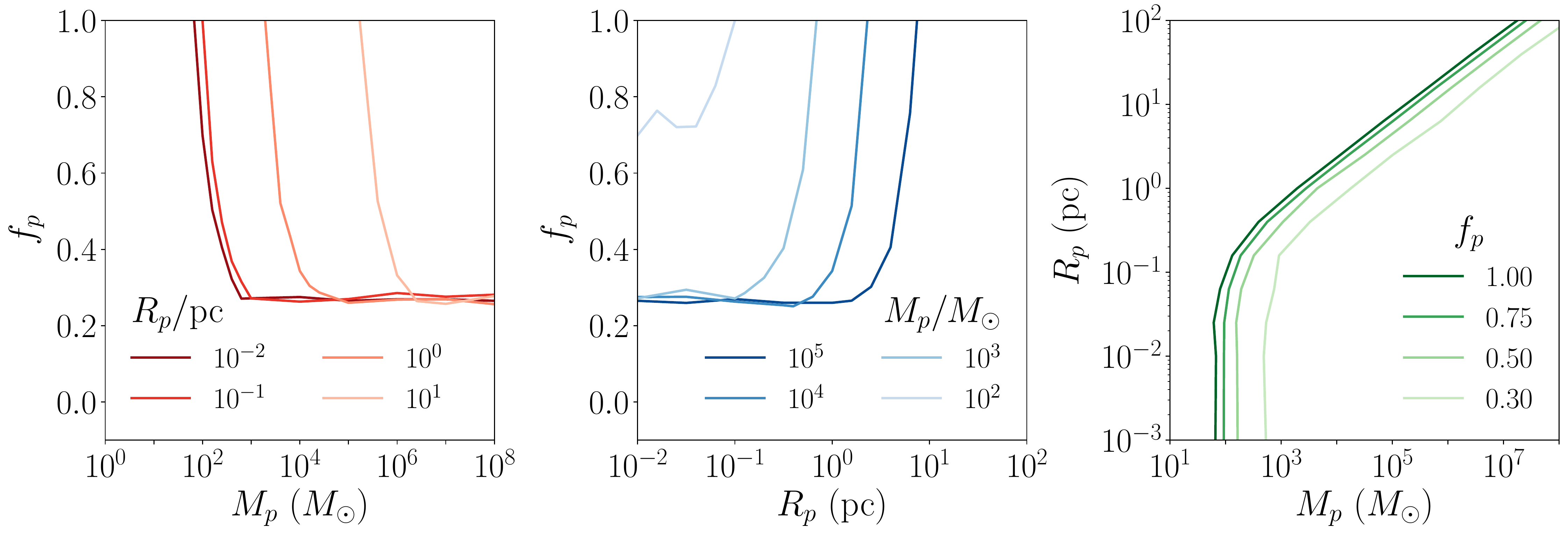}   
    \caption{Constraints on general uniform-density perturbers. \textit{(Left)} Limits on $f_{p}$ over a range of masses $M_{p}$ for discrete values of perturber radius $R_{p}$. \textit{(Middle)} Limits on $f_{p}$ over a range of radii $R_{p}$ for discrete values of perturber mass $M_{p}$. \textit{(Right)} Contours of $f_{p}$ limits in $(M_{p}, R_{p})$-space. } \label{fig:constraints_vs_uni}
\end{figure*}

In this section, we now set limits on subhalos with different total mass, radius, and density distributions. First, we continue analyzing populations of uniform-density perturbers to show how our constraints depend on the perturber mass and radius. Next, we vary the density profile along with the mass and radius by considering perturbers with power-law density profiles. Finally, we set limits on a population of Milky Way-like subhalos whose density distributions follow a Navarro-Frenk-White (NFW) density distribution \cite{navarro1996}, as predicted by $N$-body simulations. Throughout this section, we set constraints using scattering matrices calculated by simulating 5,000 binaries per bin of initial semimajor axis.

\subsection{Limits on Uniform-Density Perturbers}

To analyze how the constraints on our uniform-density 0.1~pc perturbers depend on the perturber mass $M_{p}$, we run our Monte Carlo technique and statistical analysis for several perturber populations, with masses between $10 \ M_{\odot}$ and $10^{8} \ M_{\odot}$. 
The results are shown in Fig.~\ref{fig:constraints_vs_Mp}. We find that perturbers with $M_{p} \gtrsim 95 \ M_{\odot}$ cannot make up $100\%$ of the local dark matter density at the 95\% level. Above this mass, $f_p$ can be at most $\sim 25\%$ of the local dark matter density. 

\begin{figure*}[t!] 
    \centering
    \includegraphics[width = 1.95\columnwidth]{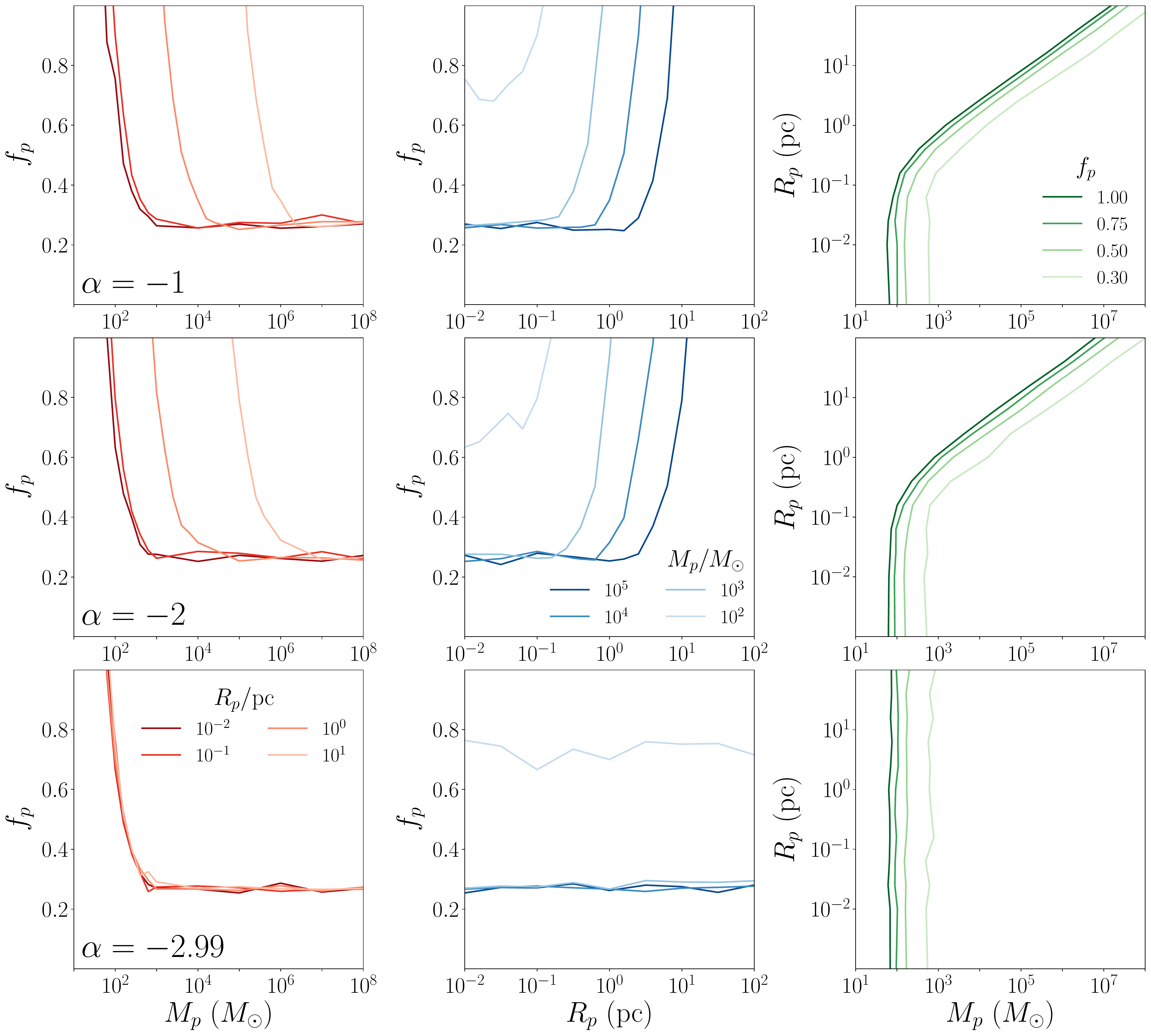}   
    \caption{Limits on perturbers with power-law density profiles. As in  Fig.~\ref{fig:constraints_vs_uni} the columns correspond to limits on $f_p$ vs.~$M_p$ for discrete values of $R_p$ (left), $f_p$ vs.~$R_p$ for discrete values of $M_p$ (center), and $R_p$ vs.~$M_p$ for discrete values of $f_p$ (right), with each row corresponding to perturbers with different power-law indices $\alpha$.} \label{fig:constraints_vs_multiple}
\end{figure*}

We next consider the limits on uniform-density perturbers as both the mass $M_p$ and radius $R_p$ are varied. The results are shown in Fig.~\ref{fig:constraints_vs_uni}. In the left panel, we show the maximum value of $f_p$ allowed by the data as a function of $M_p$ for various choices of $R_p$. In the middle panel, we show the limits as a function of $R_p$ for different values of $M_p$. As can be seen, as the radius of the perturber increases, the perturber mass at which $f_p = 1$ is ruled out increases as well; at high mass or small radius, the maximum perturber fraction asymptotes to $f_p \sim 0.25$. The right panel of Fig.~\ref{fig:constraints_vs_uni} shows the contours of the maximum $f_p$ as a function of $M_p$ and $R_p$. For radii below $\sim 0.1$~pc, the limits on $f_p$ are independent of $R_p$. For $R_p \gtrsim 0.1$~pc, the contours of constant $f_p$ behave approximately as $M_p \propto R_p^2$.

\subsection{Limits on Power-Law Perturbers}

Beyond mass and size, we expect our limits to depend on the perturber density profile. To quantify this dependence, we set limits on perturbers with various power-law density profiles, $\rho(r; \alpha) \propto r^{\alpha}$, truncated at radius $R_{p}$ and normalized to mass $M_{p}$, see Eq.~\eqref{eq:powerlaw_density_function}.

\begin{figure}[t!] 
    \centering
    \includegraphics[width = 0.9\columnwidth]{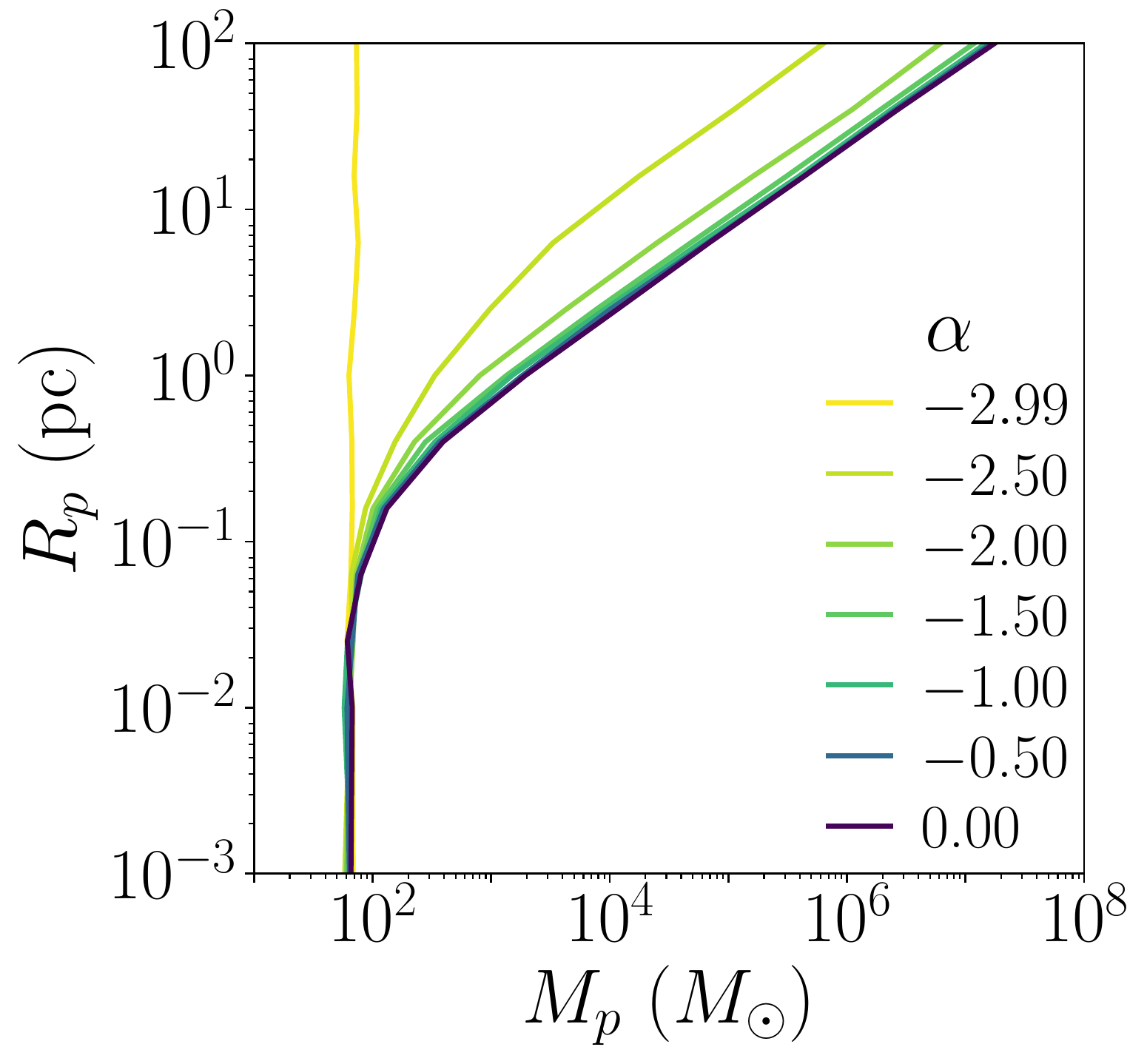}   
    \caption{$f_{p} = 1$ contours for various perturber density power-law indices $\alpha$.} \label{fig:constraints_vs_alpha}
\end{figure}

The results are shown in Fig.~\ref{fig:constraints_vs_multiple}. These plots show that constraints are generally stronger for perturbers with higher central densities. The strengthening of the limits is most significant when the power-law index $\alpha \lesssim -2$. This is most clearly seen by directly comparing the contour corresponding to $f_{p} = 1$ as density is varied, as is shown in Fig.~\ref{fig:constraints_vs_alpha}. Recall that as $\alpha \to -3$, the energy injection approaches that of a point-mass perturber, and so the dependence on $R_p$ disappears.

\subsection{Limits on NFW Perturbers}

Having considered dark matter perturbers that follow a simple power-law density distribution, we apply our formalism to constrain dark matter subhalos that follow more complicated density distributions. We consider subhalos having an NFW density profile. While other dark matter density distributions (e.g., Einasto) also exist in the literature, the NFW profile is observed to provide good fits to dark matter distributions across a wide range of halo masses both observationally \cite{Okabe_2013, schulz_10, newman_15} and in simulation \cite{springel_2008}.

The NFW distribution transitions from an $\alpha = -1$ power law for radii below the scale radius $R_s$ to $\alpha = -3$ for larger radii, before being truncated at the virial radius $R_V$:
\begin{equation}
\rho_{\rm NFW}(r;R_s,R_V) = \left\{ \begin{array}{cr} \rho_0 \left ( \frac{r}{R_s} \right )^{-1} \left( 1+\frac{r}{R_s}\right)^{-2}, & r \leq R_V \\ 0, & r > R_V, \end{array}\right.
\end{equation}
where the density parameter $\rho_{0}$ sets the virial mass $M_{V}$. In this way, the NFW profile has three free parameters: $M_V$, $R_s$, and $R_V$. 

Following the typical notation, we define the virial radius $R_{V}$ in terms of $R_s$ and a dimensionless concentration parameter $c$: $R_V \equiv c R_s$. For subhalos within a Milky Way-like host galaxy, Ref.~\cite{moline2017} used $N$-body simulations to derive the following concentration-mass relationship:
\begin{align}
c\left(M_{V}, x_{\mathrm{sub}}\right) = & c_{0}\left[1+\sum_{i=1}^{3}\left[a_{i} \log \left(\frac{M_{V}}{10^{8} \ h^{-1} \ M_{\odot}}\right)\right]^{i}\right] \times \nonumber \\
& {\left[1+b \log \left(x_{\mathrm{sub}}\right)\right] }, \label{eq:concentrationmassrelationship}
\end{align}
where $c_{0} = 19.9$, $\vec{a} = (-0.195, 0.089, 0.089)$, $b = -0.54$, and the parameter $x_{{\rm sub}}$ is the ratio between the distance of the subhalo from the center of its host halo and the host halo's virial radius. We take the former to be the Galactocentric distance to the Sun $R_{\odot} \sim 8 \ \rm{kpc}$ \cite{abuter2019, zyla2020} and the latter to be the Milky Way's virial radius $R_{V}^{\rm MW} \sim 290 \ \rm{kpc}$ \cite{deason2020}. Under these assumptions, the concentration $c$ of NFW perturbers varies between $\sim 80-120$ for subhalos with masses $\lesssim 10^8\, M_\odot$. We therefore take $c=100$ for our NFW perturbers, allowing us to quantify their density profiles with two numbers: $M_V$ and $R_V$.

We note several important caveats in the relationship given by Eq.~\eqref{eq:concentrationmassrelationship}. First, it was derived for subhalos with $R_{V} \gtrsim 10^{-1} \ {\rm pc}$, which is larger than the lower limit of perturber radii we consider. Second, the smallest simulated subhalos were evolved only to redshift $z = 32$. We will assume the concentration-mass relation does not change significantly up to $z = 0$. Third, we expect subhalos to experience tidal effects that affect their masses and density profiles. As a notable example, this relation does not account for the presence of baryonic matter. Overall, the properties of dark matter halos below $\sim 10^6\, M_\odot$ are as yet not observationally constrained. 

\begin{figure}[h!] 
    \centering
    \includegraphics[width = 0.9\columnwidth]{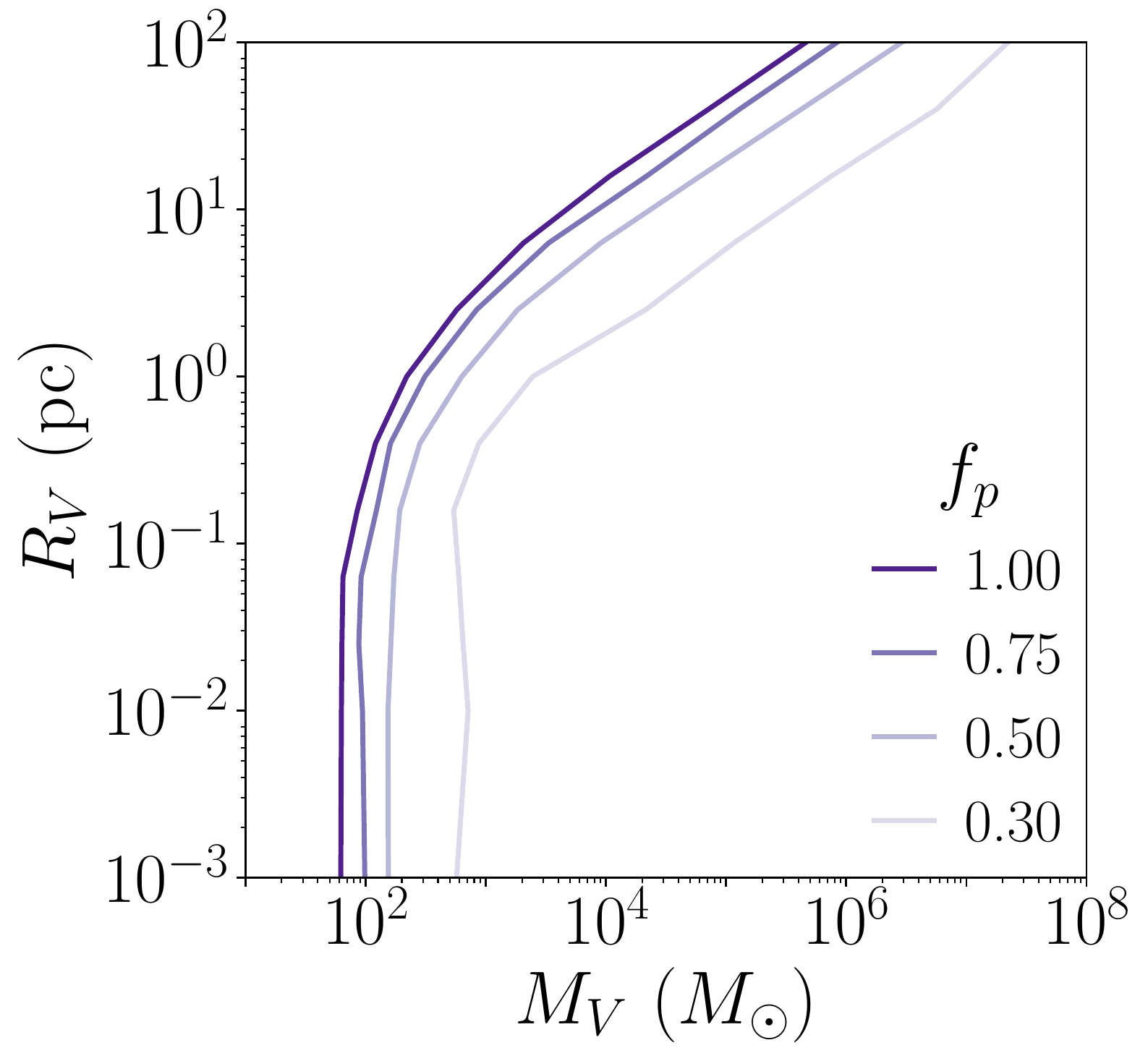}   
    \caption{Limits on NFW subhalos in the Milky Way. Here, the virial mass $M_{V}$ and the virial radius $R_{V}$ are allowed to vary, while the concentration parameter is fixed to $c=100$.} \label{fig:nfw_c_100}
\end{figure}

With these caveats stated, in Fig.~\ref{fig:nfw_c_100}, we show the upper limits (as set by the {\it Gaia} wide binary catalogue) on $f_p$ as a function of $M_V$ and $R_V$, assuming $c = 100$.

While we have treated the virial mass and the virial radius of the NFW perturbers as free parameters, the evolution of collisionless cold dark matter is expected to provide an additional relationship between the two parameters (though the precise form of this relationship depends on the environment in which they evolved). For cold dark matter evolving under the influence of gravity only, the virial radius can be set as the radius at which the dark matter density of the halo is a factor $\Delta=200$ greater than the critical density of the Universe $\rho_{c} = 2.77 \times 10^{-7} \ h^{2} \ M_{\odot} \ {\rm pc^{-3}}$ \cite{zyla2020}. Combined with our assumption of $c = 100$ for low-mass subhalos, this allows us to specify an NFW subhalo with a single parameter, $R_V$. The ``canonical'' virial mass of an NFW profile with virial radius $R_V$ we denote as $M_V^*$: 
\begin{equation}
	M_V^* = \left ( \frac{4\pi R_{V}^{3} }{3} \right ) \rho_{c} \Delta. \label{eq:MVequation}
\end{equation}

As we will show, the subhalos predicted by Eq.~\eqref{eq:concentrationmassrelationship} and Eq.~\eqref{eq:MVequation} have too little mass (for a given $R_V$) to be constrained by the wide binary data. Defining the NFW virial mass as $M_V \equiv \chi M_V^*$, we show in Fig.~\ref{fig:nfw_chi} the upper limits on $\chi$ as a function of $R_V$ (or $M_V^*$). These limits show that subhalos must be at least 5,000 times more massive than the prediction of NFW profiles from cold dark matter simulations to be constrained.

\section{Conclusions} \label{sec:conclusions}

We have constrained dark matter subhalos in a model-independent way using a catalogue of \textit{Gaia} eDR3 wide binary candidates. In general, we find that subhalos with length scales $\lesssim \ 0.1 \ {\rm pc}$ and masses $\gtrsim 65 \ M_{\odot}$ cannot make up 100\% of the dark matter (Fig. \ref{fig:constraints_vs_alpha}). The limit in the subhalo abundance drops from 100\% of the local dark matter density to around 25\% as the mass increases to $\sim 1,000 \ M_{\odot}$ (Fig.~\ref{fig:constraints_vs_multiple}). For scales $\gtrsim 0.1 \ {\rm pc}$, we found constraints to be dependent on the subhalo density profile such that higher central densities are given stronger constraints (Fig. \ref{fig:constraints_vs_alpha}). 

\begin{figure}[t!] 
    \centering
    \includegraphics[width = 0.9\columnwidth]{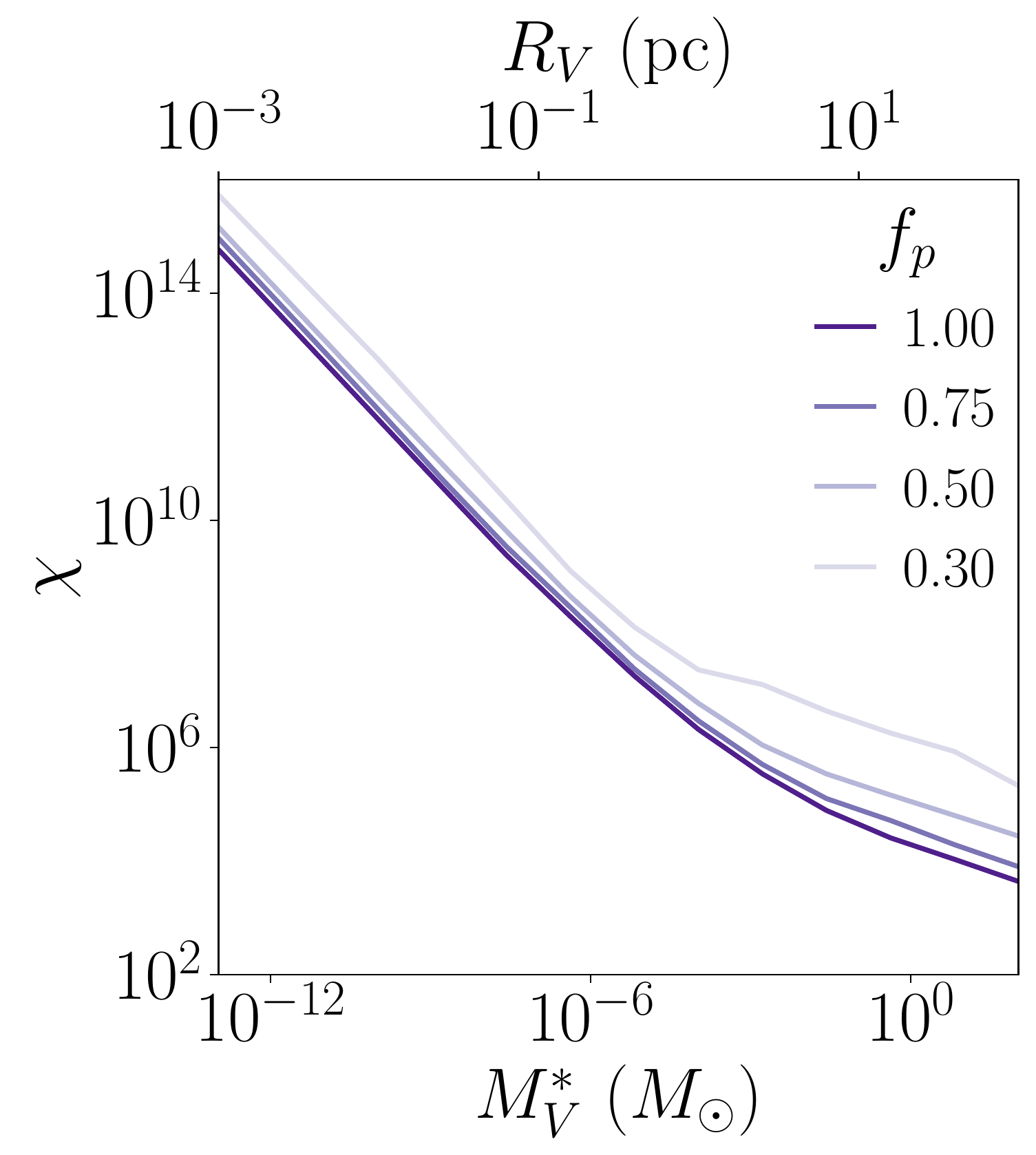}   
    \caption{Limits on modified cold dark matter NFW subhalos in the Milky Way. Here, the subhalo mass $M_{V}$ is a rescaling of its canonical virial mass: $M_{V} = \chi M_{V}^{*}$, while the rest of the NFW parameters are held fixed to their canonical values.} \label{fig:nfw_chi}
\end{figure}

In addition, we calculated how much subhalos with an NFW profile can deviate from the predictions of cold dark matter modelling without being constrained by our binary sample (Fig. \ref{fig:nfw_chi}). Across all length scales probed by our binaries, constraints apply only to subhalos that are at least 5,000 times more massive than those predicted by simulation. While not constraining collisionless cold dark matter scenarios, additional interactions within the dark sector can lead to significantly denser substructure \cite{Buckley:2017ttd,Choquette:2018lvq,bai20,Fernandez:2022zmc}. As this work sets the first limits on subhalos at $\mathcal{O}(1 \ {\rm pc})$ length scales, wide binaries can be used to constrain new regions of parameter space for dark matter models.

We have focused on constraining populations of subhalos each with a monochromatic mass spectrum. To set constraints on subhalos with extended (time-independent) mass functions, it is possible to modify the scattering formalism to include specific choices for the mass function. However, the approach of Ref.~\cite{carr2017} to extract limits on extended mass functions of primordial black holes from monochromatic constraints can be applied to our results as well.

Given our constraints $f_{p}(M_{p}) \leq f_{\max}(M_{p})$ on a monochromatic perturber mass function, one can estimate constraints on subhalos with the mass function $\psi(M_{p}) \propto M_{p} \ dn/dM_{p}$, normalized so that the fraction of dark matter existing as the subhalos is given by $f_{\psi} \equiv \int dM_{p} \ \psi(M_{p})$.  The constraint for the extended subhalo population can be estimated from the following inequality:
$$
\int dM_{p} \frac{\psi(M_{p})}{f_{\max }(M_{p})} \leq 1.
$$

Limits on extended dark matter substructure may be improved in the future in a number of ways:
\begin{enumerate}

\item Our statistical modelling allows wide freedom for the initial distribution of wide binaries. A better understanding of the binary production mechanism may restrict the viable parameter space, leading to tighter constraints on the characteristic break in the binary separation distribution due to tidal interactions with dark matter subhalos.

\item A sample of binary candidates with fewer chance alignments would reduce the uncertainty of our constraints from marginalizing over the contamination model parameters.

\item Our constraints were derived assuming a subhalo density set in terms of the local dark matter density around the Sun. Binary kinematic data allows us to better account for changes in each binary's local dark matter density as they orbit about the Galaxy \cite{quinn09,rodriguez14,tyler22}.\footnote{Varying local dark matter densities have previously been taken into account by rescaling $\rho_{\rm DM}$ to the mean time-averaged dark matter density experienced by a subsample of binaries with known velocities \cite{rodriguez14, tyler22}. Out of our wide binary catalogue, the velocities of 250 binary candidates have been measured. Following the technique of Ref.~\cite{tyler22}, we found the time-averaged local dark matter density $\langle \rho_{\rm DM} \rangle$ to have a mean of $9.7\times 10^{-3}\,M_{\odot}/{\rm pc^{3}}$ and a standard deviation of $2.9\times 10^{-3}\,M_{\odot}/{\rm pc^{3}}$.} 
Relying on \textit{Gaia} data alone, we are mainly limited by unknown radial velocities. This may improve with \textit{Gaia} DR4 \cite{evans22}, or with cross-matched data from other surveys \cite{leclerc22}.

\item Constraints can be further improved through modelling and inclusion of the various tidal effects that subhalos experience in the presence of baryonic matter.

\item A larger sample of binaries will increase the statistical power of our method.  The number of Milky Way halo/thick disk binaries available can be increased by either using larger comprehensive datasets (e.g., \textit{Gaia} DR3) or by cross-matching binaries existing in various datasets (LAMOST, APOGEE, RAVE, GALAH, GDS). Alternatively, it may be possible to survey wide binaries in ultrafaint dwarf galaxies (e.g., Draco II \cite{wilson_55}) using high-power space telescopes, such as the JWST \cite{Gardner_2006}, which opens the possibility of setting limits on substructure outside of the Milky Way \cite{2021jwst.prop.2352W}.

\end{enumerate}

\section*{Acknowledgements}
We thank Kareem El-Badry for providing a selection function for the \textit{Gaia} eDR3 dataset and Adrian Price-Whelan for assistance in correcting for the extinction using \textit{Gaia} eDR3 photometry. This work was done using the \textsc{NumPy} \cite{harris2020}, \textsc{SciPy} \cite{virtanen2019}, \textsc{AstroPy} \cite{astropy:2013, astropy:2018}, \texttt{pandas} \cite{mckinney2010}, and \texttt{corner} \cite{mackey17} packages for \textsc{Python3} and the \textsc{Mathematica} software system \cite{mathematica}.

The authors are supported by the DOE under Award Number DOE-SC0010008.

This work has made use of data from the European Space Agency (ESA) mission
{\it Gaia} (\url{https://www.cosmos.esa.int/gaia}), processed by the {\it Gaia}
Data Processing and Analysis Consortium (DPAC,
\url{https://www.cosmos.esa.int/web/gaia/dpac/consortium}). Funding for the DPAC
has been provided by national institutions, in particular the institutions
participating in the {\it Gaia} Multilateral Agreement.

\section{Data Availability}
The code we used to construct our binary sample, perform our simulations, and set constraints on dark matter subhalos is publicly available in the GitHub Repository at \url{https://github.com/edwarddramirez/dmbinaries}. 

\appendix

\begin{figure*}[t!] 
    \centering
    \includegraphics[width = 1.95\columnwidth]{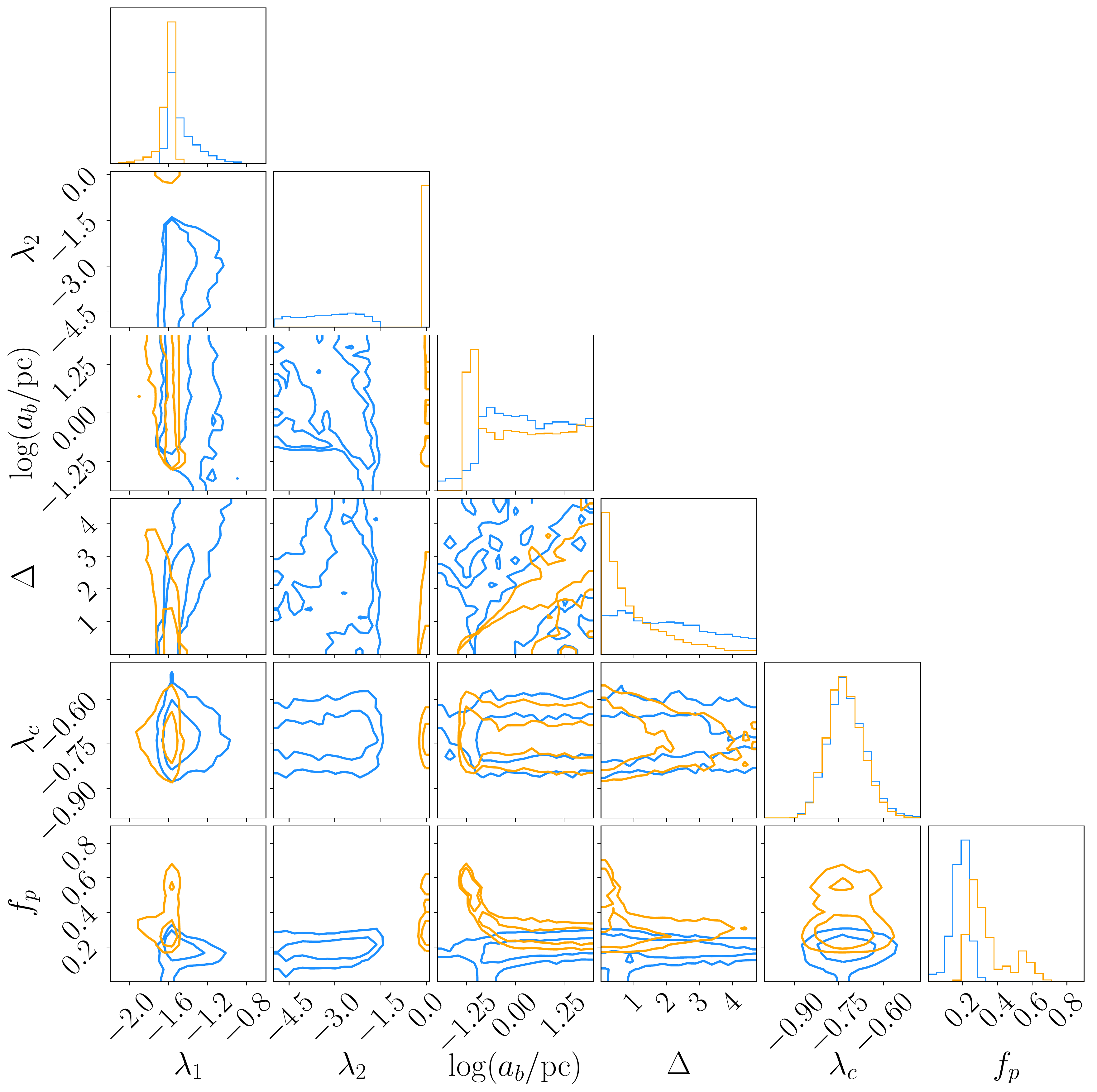}   
    \caption{Sampled posterior distributions of the model parameters $\{ \lambda_{1}, \lambda_{2}, a_{b}, \Delta, \lambda_{c}, f_{p} \}$ for uniform-density perturbers with $(M_{p}, R_{p}) = (10^{3} \ M_{\odot}, 0.1 \ {\rm pc})$ and binaries whose initial semimajor axis distribution obeys the smoothly broken power law given by Eq. \eqref{eq:sbp}. The orange and blue lines are the result of marginalizing over power-law indices satisfying $\lambda_{1}, \lambda_{2} < 0$ and $\lambda_{2} < \lambda_{1} < 0$, respectively. The inner and outer boundaries of the 2D contours denote 68\% and 95\% error contours.} \label{fig:corner-sbp}
\end{figure*}

\section{Modelling the Initial Semimajor Axis Distribution} \label{app:sbplaw}

Though the initial semimajor axis distribution of wide binaries is generally taken to obey a power law, this assumption is in part motivated by observation of the \textit{present-day} distribution \cite{tian19, lepine07, andrews_2017}. It is possible the initial distribution deviates from a simple power law for reasons that are independent of perturber interactions, due to some unidentified production mechanism or post-production assembly. T19 proposes that the initial semimajor axis distribution of wide binaries might instead be drawn from a broken power law.

To address this, we model the initial distribution of binary semimajor axes $a_{0}$ using a smoothly broken power-law distribution, which takes the following form: 
\begin{equation} \label{eq:sbp}
    \phi_{0}(a_{0}) \propto \left( \frac{a_{0}}{a_{b}} \right)^{\lambda_{1}} \left \{ \frac{1}{2} \left [ 1+\left ( \frac{a_{0}}{a_{b}} \right )^{1/\Delta} \right ] \right \}^{(\lambda_{2}-\lambda_{1})\Delta}_{\mathlarger{,} }
\end{equation}
where $a_{b}$ sets the scale at which the power law transitions from index $\lambda_{1}$ to $\lambda_{2}$ and $\Delta$ specifies the speed of the transition. As we marginalize over these parameters, we restrict ourselves to formation mechanisms that lead to a decrease in the number of binaries with respect to increasing $a_{0}$, so $\lambda_{1}, \lambda_{2} < 0$. 

\begin{figure}[b!] 
    \centering
    \includegraphics[width = 0.9\columnwidth]{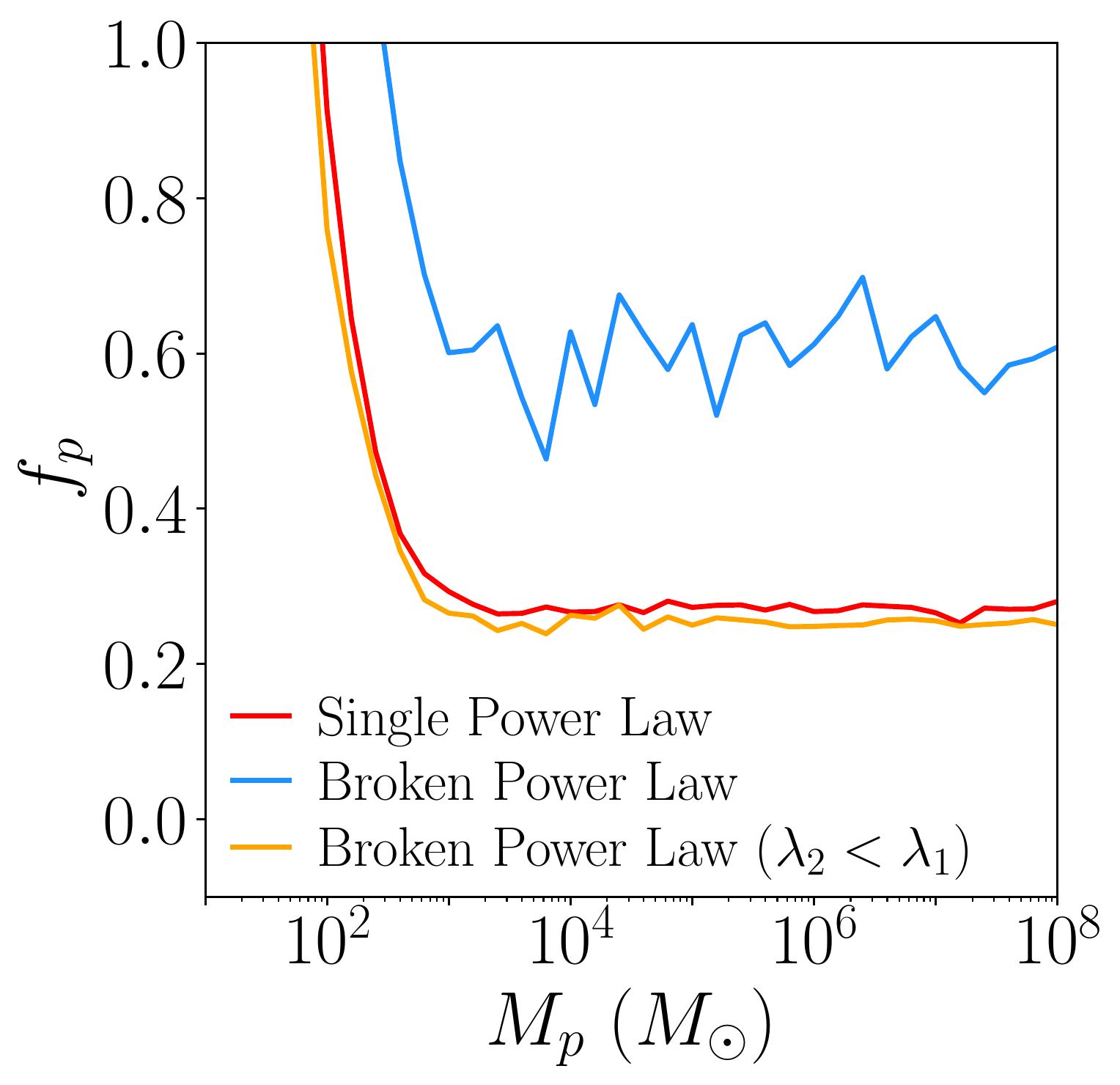}   
    \caption{Limits on populations of uniform-density perturbers with different masses $M_{p}$ and $R_{p} = 0.1~\rm{pc}$ for models in which binaries have an initial semimajor axis distribution given either by a single power law, a smoothly broken power law satisfying $\lambda_{1}, \lambda_{2} < 0$, or a smoothly broken power law satisfying $\lambda_{2} < \lambda_{1} < 0$.} \label{fig:constraints-sbp}
\end{figure}

Our updated constraints corresponding to this choice of initial semimajor axis distribution are given in Figs.~\ref{fig:corner-sbp}-\ref{fig:constraints-sbp}. As we see, our constraints are weaker, mainly due to the data preferring a model where $\lambda_{2} \sim 0$ and $a_{b} \sim 0.1$ pc. That is, the preferred fit in this case is for the widest binary assembly to be independent of semimajor axis. The {\it observed} decrease in the binary population at large $s$ then would be primarily due to encounters with the dark matter perturbers.

Though these results suggest that our constraints would be significantly weaker under the assumption of a broken power-law production mechanism, it is plausible that the assembly process for the widest binaries should be less efficient as the semimajor axis increases (restricting the possible values of $\lambda_2$ and $\lambda_1$). For instance, it has been proposed that wide binaries with separations $\gtrsim 0.1 \ {\rm pc}$ were most likely formed as a random alignment of two stars with low relative velocities in an expanding cluster \cite{tyler22, moeckel_2011, moeckel_2010, kouwenhoven_10, griffiths18}. This ``soft capture" occurs on timescales of 20 - 50 Myr for each cluster \cite{kouwenhoven_10}, much shorter than the 10 Gyr evolution time within the halo and thick disk that this paper is concerned with. The distributions of inter-star distances within the cluster result in a falling distribution of binary semimajor axes under this mechanism.

As argued in T19, wide binaries produced from a single cluster would have an initial semimajor axis distribution given by the power law $\phi_{0}(a_{0}) \propto a_{0}^{-3/2}$, breaking at a characteristic length scale (corresponding to the cluster's tidal radius) to a steeper decline of wide binaries. For binaries formed within many different clusters, the overall distribution of semimajor axes would be the combination of various $a_0^{-3/2}$ power laws, each having breaks at different scales. This results in a distribution of binary semimajor axes that behaves as $a_{0}^{-3/2}$ for small $a_{0}$ and eventually breaks to a more rapidly decreasing distribution at large $a_{0}$ in a way that depends on properties of the cluster population. Approximating the large-$a_0$ distribution as a power law, this implies the index $\lambda_2$ at large $a_0$ is strictly less than the index at small $a_0$, $\lambda_1$.

Our results requiring that $\lambda_2 < \lambda_1$ are shown in  Figs.~\ref{fig:corner-sbp}-\ref{fig:constraints-sbp}. In this case, the constraints are nearly identical to those set using the single power law assumption for the initial semimajor axis distribution.  Thus, our constraints are robust under the assumption that the initial distribution of wide binaries is decreasing, with the number of binaries at high separations decreasing as fast as or faster than that at low separations. 

\section{Chance-Alignment Modelling} \label{app:chance-align}
We have set constraints on subhalos assuming that the distribution of chance alignments (subject to the various quality cuts used to construct the catalogue) follows a power law as a function of projected separation $s$. In this Appendix, we show that our results agree with constraints that are set assuming two other functional forms for the chance-alignment separation distribution.

\begin{figure}[b!] 
    \centering
    \includegraphics[width = 0.9\columnwidth]{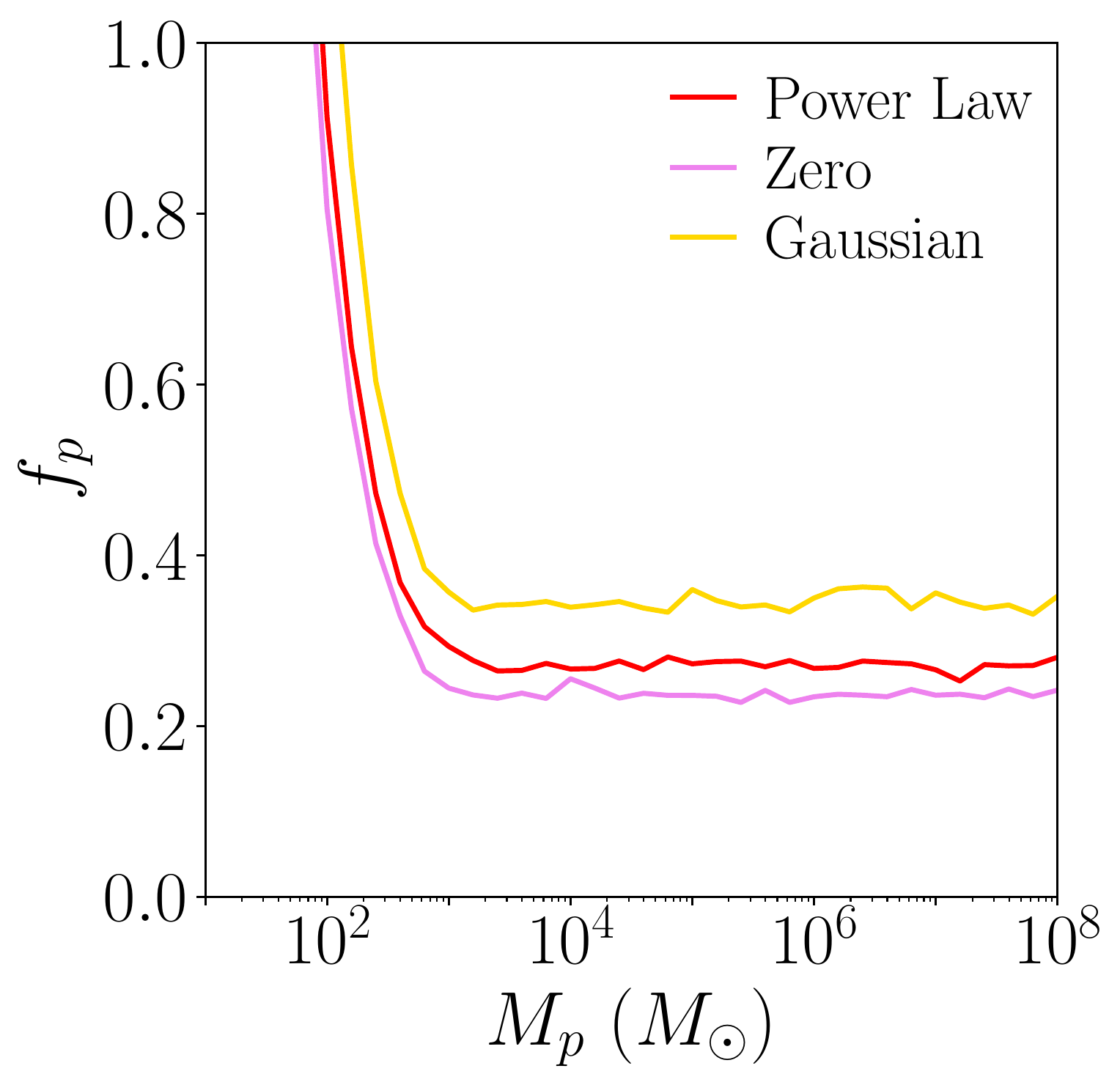}   
    \caption{Limits on populations of uniform-density perturbers with different masses $M_{p}$ and $R_{p} = 0.1~\rm{pc}$ for models in which the projected separation distribution of chance alignments is either a single power law, identically zero, or a Gaussian.} \label{fig:ca}
\end{figure}

\begin{figure}[t!] 
    \centering
    \includegraphics[width = 0.9\columnwidth]{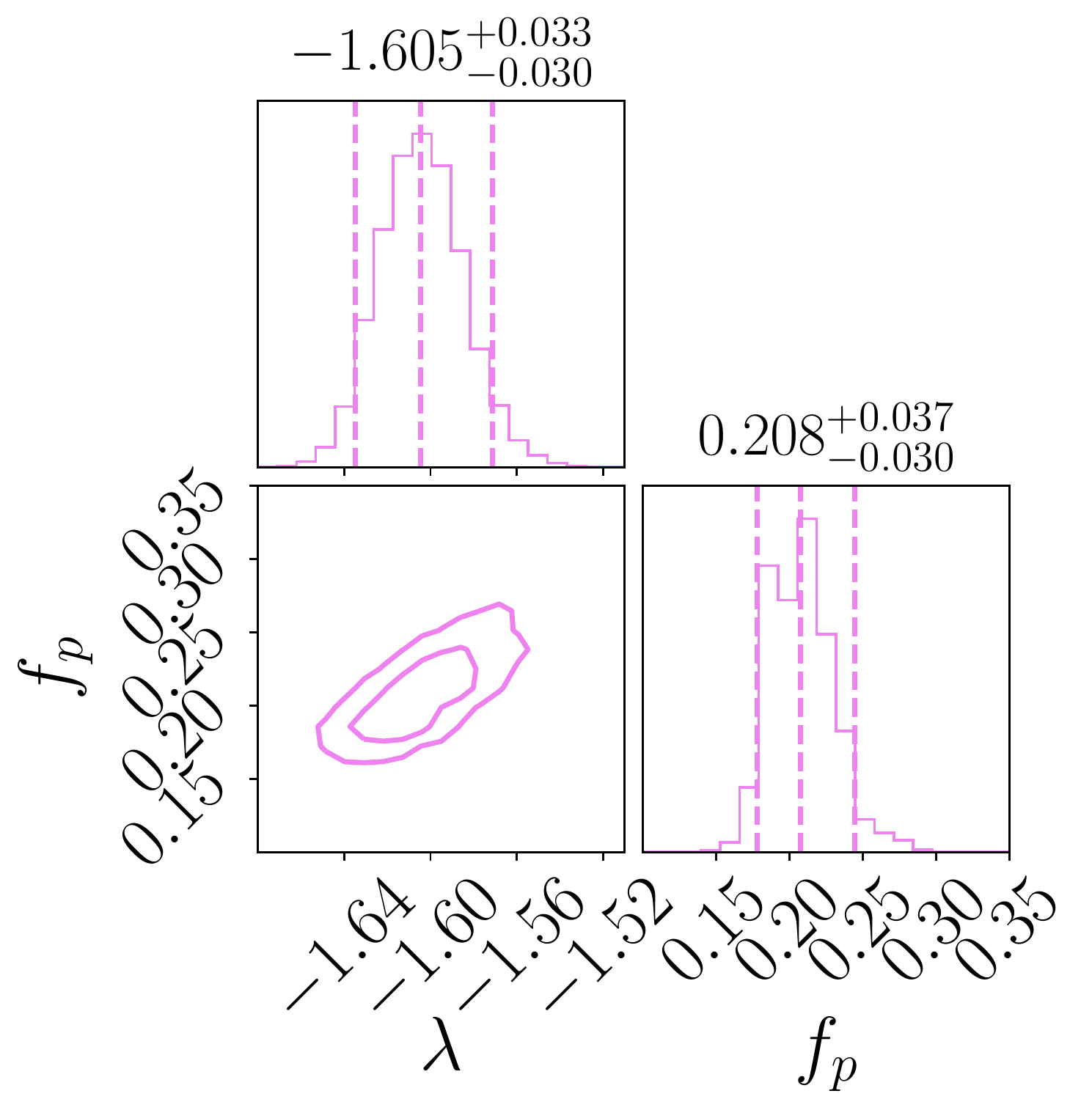}   
    \caption{Sampled posterior distribution of the model parameters $\{ \lambda, f_{p} \}$ for uniform-density perturbers with~$(M_{p}, R_{p}) = (10^{3} \ M_{\odot}, 0.1 \ {\rm pc})$ and no chance-alignment model $(\phi_{c} = 0)$. The vertical dashed lines in the 1D histograms denote 5\%, 50\%, and 95\% quantiles. The inner and outer boundaries of the 2D contours denote 68\% and 95\% error contours.} \label{fig:noca}
\end{figure}

First, we set constraints without taking the population of chance alignments into account. This corresponds to setting the chance-alignment distribution $\phi_{c} = 0$ in Eq.~\eqref{eq:probability_final}. As we see from Fig.~\ref{fig:ca}, minimizing the effect of chance alignments in this way does not significantly alter our constraints. The posterior corresponding to 0.1 pc uniform-density perturbers with $M_{p} = 10^{3} \ M_{\odot}$ is given in Fig.~\ref{fig:noca}.

\begin{figure}[t!] 
    \centering
    \includegraphics[width = 0.9\columnwidth]{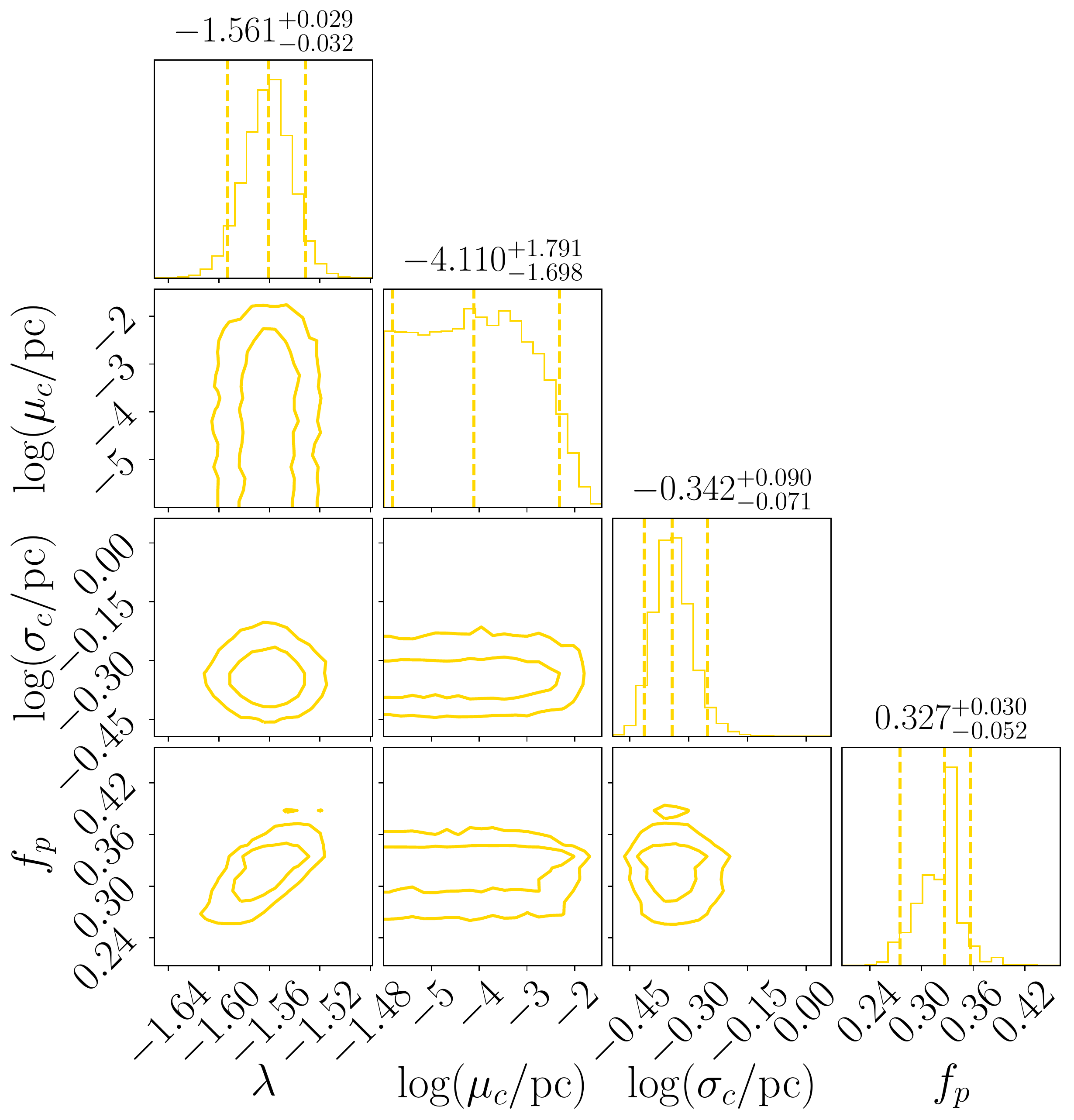}   
    \caption{Sampled posterior distribution of the model parameters $\{ \lambda, \mu_{c}, \sigma_{c}, f_{p} \}$ for uniform-density perturbers with $(M_{p}, R_{p}) = (10^{3} \ M_{\odot}, 0.1 \ {\rm pc})$ and a chance-alignment separation distribution given by a Gaussian, see Eq.~\eqref{eq:gaussian}. The vertical dashed lines in the 1D histograms denote 5\%, 50\%, and 95\% quantiles. The inner and outer boundaries of the 2D contours denote 68\% and 95\% error contours.} \label{fig:gca}
\end{figure}

Next, we consider a Gaussian chance-alignment distribution:
\begin{equation} \label{eq:gaussian}
    \phi_{c}(s) = \frac{1}{\sqrt{2\pi \sigma_{c}^{2}}} \exp \left [ -\frac{1}{2} \left ( \frac{s - \mu_{c}}{\sigma_{c}} \right )^{2} \right ],
\end{equation} 
where $\mu_{c}$ and $\sigma_{c}$ denote the mean and standard deviation, respectively. The corresponding limits are given in Fig.~\ref{fig:ca}; they are consistent with limits from the default single power law and as well as the no-chance-alignment limits. The posterior corresponding to 0.1 pc uniform-density perturbers with $M_{p} = 10^{3} \ M_{\odot}$ is given in Fig.~\ref{fig:gca}.

\bibliography{references}  

\end{document}